\documentclass[11pt]{article}

\usepackage[
    tmargin=2.5cm,
    bmargin=2.5cm,
    lmargin=2.5cm,
    rmargin=2.5cm
]{geometry}

\usepackage{amsmath}
\usepackage{amssymb}
\usepackage{amsfonts}
\usepackage{amsthm}
\usepackage{bbm}
\usepackage{bm}
\usepackage{changepage}
\usepackage{tikz}
\usetikzlibrary{arrows.meta,calc,positioning}
\definecolor{htmlblue}{HTML}{0000FF}
\definecolor{othergray}{HTML}{888888}
\definecolor{edgegray}{HTML}{4A4A4A}

\usepackage{graphicx}
\usepackage{booktabs}
\usepackage{array}
\usepackage{adjustbox}
\usepackage{threeparttable}
\usepackage{subcaption}
\usepackage{pdflscape}
\usepackage{comment}
\usepackage{titletoc}

\usepackage{floatrow}

\usepackage[
    justification=justified,
    singlelinecheck=false
]{caption}

\usepackage{setspace}
\usepackage{enumitem}
\usepackage{ragged2e}
\usepackage{ulem}
\usepackage{comment}
\usepackage{footmisc}
\usepackage{xcolor}
\usepackage{eurosym}
\usepackage{fancyhdr}
\usepackage{tabularx}

\usepackage[authoryear,round]{natbib}

\setcitestyle{
    aysep={},
    yysep={,},
    citesep={;},
    notesep={, }
}
\usepackage{xr-hyper}

\usepackage{datetime}
\newdateformat{monthyeardate}{%
    \monthname[\THEMONTH] \THEYEAR
}

\usepackage[
    bookmarks=true,
    bookmarksnumbered=true,
    bookmarksopen=false,
    breaklinks=true,
    hypertexnames=false,
    colorlinks=true,
    citecolor=blue,
    linkcolor=blue,
    urlcolor=blue
]{hyperref}
\hypersetup{
    pdftitle={
       The Evolution of Technology Portfolios and Firm Profitability in the European Power Sector
    },
    pdfauthor={Robin Fischer and Anton Pichler}
}

\allowdisplaybreaks

\newcommand{\email}[1]{\href{mailto:#1}{\textcolor{black}{\texttt{#1}}}}

\makeatletter

\newcommand*{\addFileDependency}[1]{%
    \typeout{(#1)}%
    \@addtofilelist{#1}%
    \IfFileExists{#1}{}{%
        \typeout{No file #1.}%
    }%
}

\makeatother

\theoremstyle{plain}

\newcolumntype{R}[1]{>{\RaggedRight\arraybackslash\hsize=#1\hsize}X}

\graphicspath{{figures/}{./}}

\begin{document}


\title{
    The Evolution of Technology Portfolios and Firm Profitability \\ in the European Power Sector\\[0.4em]
}



\author{%
    \vspace{1.5cm}
}

\author{%
    Robin Fischer
    \and
    Anton Pichler%
    \thanks{%
        Corresponding author: Anton Pichler,
        Vienna University of Economics and Business;
        Complexity Science Hub Vienna; and Macrocosm, USA,
        \email{anton.pichler@wu.ac.at}.
        Robin Fischer:
        Vienna University of Economics and Business; Complexity Science Hub Vienna,
        \email{robincfischer@gmail.com}.
        We thank Jan Hurt, Rupert Way, Kavita Surana, and Christian Rumpelnik
        for helpful comments and discussions.
        Generative AI tools were used in the preparation of this
        manuscript. OpenAI Codex (GPT-5) was used to support language
        editing and replication-code auditing. The authors reviewed
        and verified all AI-assisted output and take full responsibility
        for the content of the manuscript.
    }
}

\date{%
    \vspace{0.25cm}
    October 2026
}

\maketitle

\vspace{0.25cm}


\renewenvironment{abstract}{%
  \par
  \begin{center}
    \begin{minipage}{0.85\textwidth}
      \begin{center}
        \normalsize\bfseries Abstract
      \end{center}
      \vspace{0.5em}
      \normalsize
      \setstretch{1.15}
      \setlength{\parindent}{0pt}
      \ignorespaces
}{%
      \par
    \end{minipage}
  \end{center}
  \vspace{1.5em}
}

\begin{abstract}{
Power firms' generation portfolios are long-lived and change slowly. We study how this inertia shapes profitability as market conditions shift. Linking plant-level data to financial accounts for over 1,300 EU power firms over 2002--2023, we find renewable- and fossil-oriented firms earned similar returns until 2018. Thereafter, renewable-oriented firms earned about 3.2 percentage points higher return on assets. Exposure to rising carbon and gas
prices accounts for roughly 60\% of this divergence. Durable portfolios thus turn market shifts into sustained profit differences.
}

\end{abstract}
\vfill
\thispagestyle{empty}

\newpage


\setcounter{page}{1}

\section{Introduction}
\label{sec:introduction}

The European electricity sector is undergoing a profound structural transformation.
Wind and solar, which supplied less than 1\% of electricity
generation in the European Union in 2000, accounted for 27\% by 2023 \citep{ember2024ger}.
In 2025, their combined output exceeded fossil generation
for the first time \citep{ember2026ger}. 
This transition is carried out by firms, with independent power producers and utilities owning the majority of Europe's generation assets \citep{KELSEY201865,eberhard2017accelerating}. 
Their decisions to invest in, maintain, or retire plants determine how fast, and at what cost, the generation mix changes.

Profitability considerations are central to these decisions. Retained earnings can finance new investments, and expected returns drive investment choices.
However, generation portfolios are highly persistent. 
Firms operate plants for decades, and their generation portfolios
adjust only slowly and path-dependently \citep{alova2020global,pichler2026five}. Thus, a firm's technology mix reflects investment decisions made under market conditions that may no longer exist. 
Power producers differ markedly in their past investment choices: many firms are specialized in specific technologies and some hold mixed portfolios.
When technology costs, fuel prices, and regulation change,
firms with different portfolios are exposed differently, and past
investment choices can shape profitability for many years
\citep{dixit1994investment}.
These differences matter for the energy transition because they can shape firms' incentives and ability to reallocate capital across generation technologies.

Several developments over the past decade changed the economic conditions facing these portfolios. Reforms to the EU Emissions Trading System (ETS) tightened the supply of emission allowances and were followed by a sustained increase in carbon prices \citep{eu2018etsdirective,sitarz2024eu}. The European energy crisis of 2021--22 then brought a sharp increase in natural-gas prices \citep{iea2022gasmarket}. Meanwhile, the costs of wind and solar technologies have fallen dramatically and consistently over recent decades \citep{way2022empirically}, strengthening their competitive position relative to conventional generation technologies. These changes affected firms differently depending on the technologies in their generation portfolios.

The implications of these market dynamics for firms' profitability are not obvious. For fossil generators, higher allowance prices raise production costs, but a substantial share of these costs may be passed through to wholesale electricity prices \citep{fabra2014}. Higher gas prices similarly increase the costs of gas-fired generation, but because gas plants often set the marginal electricity price \citep{zakeri2023role}, they also lift wholesale prices, and gas plants benefit from their flexibility during periods of high demand and tight supply. Renewable generators, by contrast, require no fuel inputs and incur no direct carbon costs and, thus, earn larger inframarginal rents when these costs push up wholesale prices.
But renewables are capital intensive and face ``cannibalization effects,'' whereby their market value declines as renewable penetration increases \citep{HIRTH2013218,Hirth_2016,LOPEZPROL2020104552}. The net effect of changing market conditions on the profitability of firms with heterogeneous generation portfolios is therefore ambiguous ex ante and ultimately an empirical question.

Existing empirical evidence provides little guidance on how such shifts have affected firm profitability.
Firm-level studies report negative, insignificant, and positive associations between renewable orientation and profitability \citep{RUGGIERO20173676,tomczak2019,DORIGONI2024114323}. These studies typically estimate a single average relationship over their observation period.
If the returns to different generation portfolios vary with technology costs, input prices, and regulation, however, such averages will obscure how relative profitability evolves over time.

We study how the profitability of European power firms with different generation portfolios has evolved over two decades. We link plant-level information on generation technology, capacity, and ownership to financial accounts for 1,357 power firms in the EU-27 over 2002--2023. For each firm and year, we reconstruct its installed generation portfolio across eight technologies and relate portfolio composition to realized accounting profitability, allowing this relationship to vary over time. Unlike most previous studies, our sample extends beyond listed utilities to include privately held and state-owned power producers.

We first show that generation portfolios are a highly persistent source of firm heterogeneity. Firms' dominant technology rarely changes over the two decades we observe, and firms organized around different technologies follow distinct long-run profitability trajectories. Wind-oriented power firms exhibit the strongest and most robust improvement in profitability over time, while profitability trajectories for other technologies are less distinct.

We next show that the relative profitability of renewable- and fossil-oriented firms changes markedly around 2018. Until then, the two groups earn statistically indistinguishable returns on assets, conditional on lagged firm characteristics and country-by-year fixed effects. Over 2019--2023, however, a profitability gap between the two groups arises, with renewable-oriented firms earning about 3.2 percentage points more.

Finally, we examine whether this profit divergence is associated with firms' differential exposure to changes in carbon and natural-gas prices. We find that higher allowance prices are indeed associated with lower relative profitability among firms with more carbon-intensive generation portfolios, and this association persists through 2023. By contrast, the association with gas exposure is concentrated in the 2022 crisis year. Together, carbon and gas prices account for about 60\% of the post-2018 divergence in relative profitability.

These findings contribute to three strands of research. First, we add to the literature on persistent heterogeneity among firms. We provide evidence that the inertia of generation portfolios has converted shifts in technology costs, input prices, and regulation into sustained differences in profitability in the European power sector.

Second, we contribute to empirical work on renewable orientation and financial performance. Existing studies rely largely on listed utilities, restricted geographic samples, and shorter observation periods. Using a broader population of firms, we show that the profitability of renewable- and fossil-oriented firms has differed little for most of the sample period, but a pronounced gap opens after 2018. This temporal pattern offers one explanation for the conflicting conclusions in the literature.

Third, we contribute to research on the economic effects of carbon pricing. Studies find limited adverse effects of the EU ETS on the average regulated firm. We show that these average effects conceal substantial heterogeneity within the power sector: how carbon and gas prices affect profitability depends heavily on the specific composition of firms' technology portfolios.

Taken together, our findings reveal an emerging divergence in the economic position of European power firms. Recent market conditions have increased the relative profitability of renewable-oriented portfolios, strengthening incentives to allocate capital toward renewable technologies. 
However, the extent to which these market forces translate into actual portfolio reallocation depends on how quickly firms can adjust their existing generation assets and on the regulatory and institutional constraints governing the electricity sector.

The remainder of the paper is organized as follows. Section~\ref{sec:literature} reviews the related literature, documents the development of carbon and gas prices, and describes the data. Section~\ref{sec:strategy} presents the construction of generation portfolios and the empirical framework. Section~\ref{sec:results} reports the results, and Section~\ref{sec:conclusion} concludes.

\section{Background and Data}
\label{sec:literature}

\subsection{Related Literature}
Our paper relates to work on persistent firm heterogeneity, on renewable energy and firm performance, and on the profitability effects of carbon prices.

Differences in profitability and productivity across firms are large and can persist over long periods \citep{mueller1986profits,geroski1988persistence,syverson2011determines, bottazzi2008productivity}, as firm-specific routines and capabilities change only slowly \citep{nelson1982evolutionary, dosi2017footprint}. 
Physical capital can introduce an additional source of persistence. Assets are often long-lived, and investment is partly irreversible, so past investment choices can constrain firms' current positions \citep{ghemawat1991commitment,
dixit1994investment}. In energy systems, such constraints can contribute to carbon lock-in \citep{unruh2000understanding}. 
Power generation is a case in point: plants operate for decades, and generation portfolios adjust slowly and along path-dependent trajectories \citep{alova2020global,pichler2026five}. We build on this literature by examining how persistent differences in physical asset portfolios interact with a changing market environment to shape differences in firm profitability.

A second strand of literature studies how renewable energy investments affect firm performance. Overall, the empirical evidence remains inconclusive. \citet{RUGGIERO20173676} find that the volume of renewable energy produced is negatively associated with return on assets, return on equity and Tobin's $Q$ across 66 listed electric utilities. \citet{tomczak2019} compares renewable and fossil-based generators in the Baltic States and Central Europe and finds no statistical difference in financial standing in most of the indicators examined. \citet{DOPIERALA202211492} likewise report only small differences between producer types in the Baltic Sea Region, although their descriptive series show the average ROE of wind producers rising after 2016 while that of fossil producers declines. Evidence based on stock returns and broader sets of financial indicators is similarly inconclusive \citep{chebotareva2018leading, colombelli2020green}.

The study most closely related to ours is \citet{DORIGONI2024114323}, who examine how renewable generation shares are associated with the profitability and market valuation of European energy firms. They find that higher renewable generation shares are associated with higher profitability and market valuations, but not with a lower cost of capital, and that firms with renewable shares above 80\% exhibit higher returns on assets and equity than firms with more mixed portfolios. We extend their analysis in three main respects. First, our asset-level ownership data allow us to cover a substantially larger population of European power firms over a longer period, including unlisted independent power producers and state-owned firms. 
Second, rather than considering renewable exposure in isolation, we compare renewable- and fossil-oriented generation portfolios directly. Third, we allow the profitability difference between these portfolios to vary over time.

The third strand evaluates how the EU ETS affects regulated firms' economic performance. Synthesizing early empirical work, \citet{joltreau2019does} conclude that the EU ETS had no detectable negative impact on firm profitability or competitiveness across industries during its first two phases.
More recent quasi-experimental studies reach similar conclusions.
\citet{DECHEZLEPRETRE2023102758} find that the EU ETS reduced emissions without lowering profits or employment and report increases in revenues and fixed assets. \citet{colmer2025does} find substantial emissions reductions among regulated manufacturing firms without detectable contractions in output or employment. Two main features of the early EU ETS help explain these results. Free allocation limited firms' direct expenditures on allowances, and over-allocation contributed to low allowance prices during much of the first two trading phases \citep{joltreau2019does}. 

For power producers, cost pass-through provides an additional channel.
Because allowances carry an opportunity cost, generators may incorporate their cost into electricity-market bids even when the allowances were allocated free of charge. Using detailed bidding data from the Spanish electricity market, \citet{fabra2014} estimate that approximately 80 percent of emissions costs were passed through to wholesale electricity prices.
Pass-through can offset part of the compliance costs borne by fossil generators and, under free allocation, may generate windfall profits. Low-carbon generators may benefit from the same increase in electricity prices without incurring corresponding direct allowance costs \citep{lise2010,sijm2006}. Carbon pricing may therefore have little effect on average power-sector profitability while changing the relative profitability of firms with different generation portfolios.

The empirical evidence suggests overall positive net returns of the early EU ETS for power producers. \citet{CHAN20131056} find that the introduction of the EU ETS increased unit material costs in the power sector, but was accompanied by a considerably larger increase in revenues. Evidence from financial markets points in the same direction: \citet{bushnell2013} show that the sharp fall in the allowance price in 2006 reduced the market value of firms in carbon- and electricity-intensive industries, consistent with investors expecting at least some firms to benefit from the regulation.

Closest to our price-exposure analysis is \citet{veith2009}, who examine how the stock returns of major European power companies respond to allowance-price movements. Their results indicate an overall positive association between allowance and stock returns, but also suggest that the response differs with firms' production portfolios. In particular, firms that relied more heavily on fossil generation were perceived to face an additional burden during the second trading phase. Their analysis provides early evidence that the incidence of carbon pricing may vary within the power sector. It is nevertheless based on a small sample of listed companies, considers stock-market valuations rather than realized accounting profitability, and predates the substantial increase in allowance prices after 2018.


\subsection{Carbon and Gas Price Developments}
\label{subsec:price_background}

\begin{figure}[htbp]
\centering
\includegraphics[width=\textwidth]{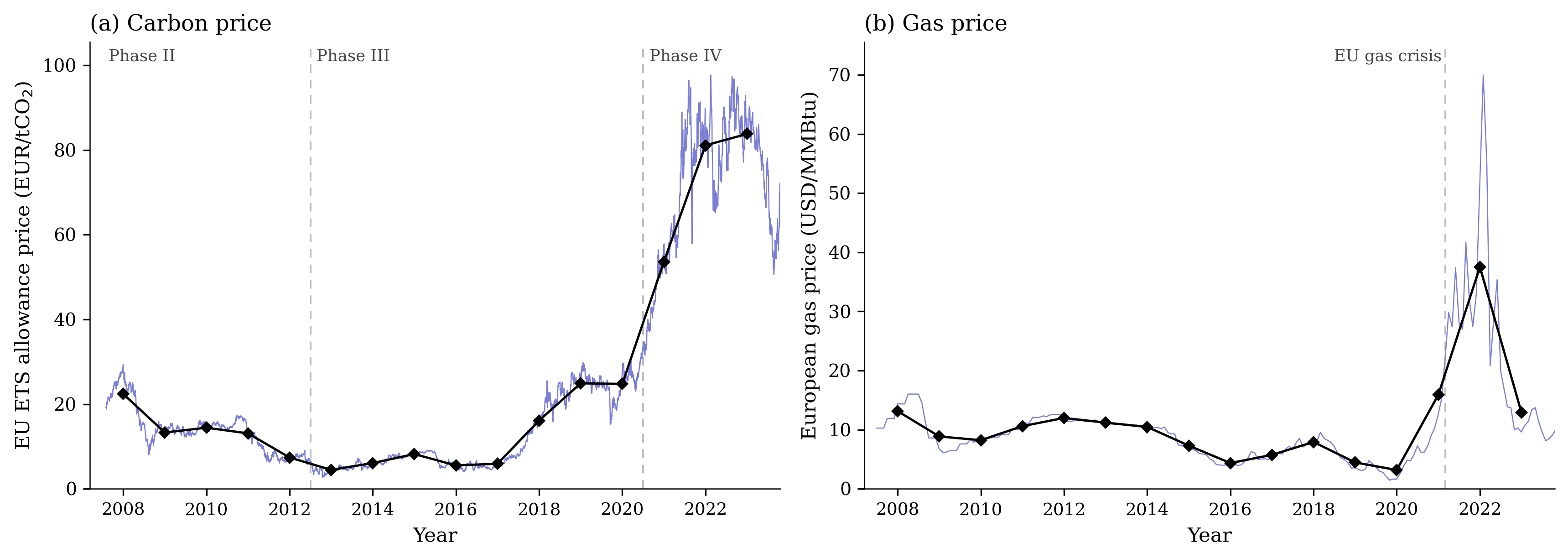}
\caption[EU ETS allowance and European natural-gas prices]{%
{EU ETS allowance and European natural-gas prices.}
Panel~(a) reports the daily ICE settlement price of EU ETS allowances (thin
line) and the corresponding calendar-year means (markers) from the beginning of
Phase~II. Dashed vertical lines mark the beginnings of Phases~III and IV.
Panel~(b) reports the monthly European natural-gas benchmark price (thin line)
and the corresponding calendar-year means (markers). The dashed vertical line
marks the onset of the European gas crisis in 2021.
\textit{Sources:} ICE; IMF via FRED (\texttt{PNGASEUUSDM}).%
}
\label{fig:prices}
\end{figure}

Carbon and natural-gas prices are two key components of the market environment faced by European power producers. Under the EU ETS, generators must surrender allowances for their direct carbon emissions, making allowance prices a component of the production costs of emitting units. Natural gas is both a major input into electricity generation and an important determinant of wholesale power prices in many European markets. Changes in the two prices can therefore affect firms' costs and revenues through distinct channels.
Figure~\ref{fig:prices} documents the development of both price series over the sample period. Allowance prices declined during the early years of Phase~II and remained comparatively low through much of Phase~III. They began to rise in 2018, reached annual means of approximately EUR~25 per tonne in 2019 and 2020, and increased further during Phase~IV, exceeding EUR~80 per tonne on average in 2022 and 2023.

European gas prices remained comparatively moderate through 2020 before rising sharply during the European energy crisis, reaching their sample maximum in 2022 and declining again in 2023. The concurrent late-sample increase in allowance and gas prices therefore marks a substantial change in two key dimensions of the economic environment faced by European power producers. Although the two prices affect firms through different channels, their sharp increases overlap in time and are therefore relevant to the same period of changing profitability conditions. We consequently account for carbon- and gas-price exposure jointly in the empirical analysis.

\subsection{Data}

Our analysis combines firm financial statements with asset-level generation data, both from S\&P Capital IQ. The financial data provide total assets, total debt, total equity, total revenue and net income, from which we construct return on assets, return on equity, leverage, firm size and sales growth; Appendix Table~\ref{app:variables} gives the exact definitions. Generation assets come from the S\&P Global Power Plant Database, with unit-level information on technology, nameplate
capacity, commissioning and retirement dates, and ownership. Because both sources share an entity identifier, financial accounts are matched to asset owners directly.

The matched panel covers 1,357 firms and 11,971 firm-years in the EU-27 between 2002 and 2023. It retains firms in four industry categories: Electric Utilities, Independent Power Producers and Energy Traders, Renewable Electricity, and Multi-Utilities. Firms in the panel owned 403 GW of generating capacity in 2023, against an EU-27 total of 1,111 GW, so the panel accounts for roughly 36 per cent of installed capacity.\footnote{Eurostat \texttt{nrg\_inf\_epc}, net maximum electrical capacity, all operators, 2023. Our figure is nameplate capacity from the plant database and is therefore
not exactly comparable; the share should be read as an approximation.} The EU ETS allowance
price is the calendar-year mean of daily ICE settlement prices and the natural-gas price the calendar-year mean of the European spot index from FRED. Observations with missing values are dropped row-wise during variable construction, and Appendix Table~\ref{tab:funnel} traces the panel from the matched data to each estimation sample.

\section{Empirical Strategy}
\label{sec:strategy}

\subsection{Measuring Technology Portfolios}
\label{subsec:portfolios}

For firm $i$ in year $t$, we observe installed capacity in megawatts (MW) across eight generation technologies,
$m \in \{\text{wind},\ \text{solar PV},\ \text{hydropower},\ \text{nuclear},\ \text{gas},\ \text{coal},\ \text{oil},\ \text{biomass}\}$. We define firm $i$'s portfolio share in technology $m$ as
\begin{equation}
s_{imt}
=
\frac{\text{MW}_{imt}}{\sum_{m'} \text{MW}_{im't}},
\label{eq:shares}
\end{equation}
such that $\sum_m s_{imt}=1$ by construction. A generating unit contributes to a firm's portfolio in year $t$ if it entered service in or before $t$ and had not been retired by that year. The resulting measure therefore represents the firm's installed generation stock rather than its flow of new investment.

We follow \citet{alova2020global} and assign generating assets to their direct owners rather than aggregating them to ultimate parent companies.\footnote{Appendix Tables~\ref{tab:gap_rob_parents}--\ref{tab:gap_rob_subs} show that the estimated post-2018 profitability gap remains statistically significant when parent companies or subsidiaries are excluded separately.}
This allows us to identify subsidiaries, independent power producers, and privately held firms that are not represented in conventional listed-firm datasets. We exclude firm-years with missing and negative capacity records. Appendix~\ref{app:data} describes the ownership assignment, portfolio construction, and data-quality checks in detail.

\subsection{Outcomes, Control Variables, and Model Uncertainty}
\label{subsec:variables}

We examine how the profitability of European power firms varies with the
composition of their generation portfolios and with their exposure to carbon
and gas prices. The underlying panel covers the period from 2002 to 2023,
although individual specifications use shorter samples where lagged variables
or additional data requirements apply.

Existing studies differ substantially in their outcome variables, definitions
of renewable orientation, sample composition, control sets, and estimation
procedures. Renewable orientation, for example, has been measured as the annual
volume of renewable generation in levels \citep{RUGGIERO20173676}, as the share
of renewable output in a firm's total generation
\citep{DORIGONI2024114323}, and as an indicator for firms drawing more than
half their output from a single technology \citep{DOPIERALA202211492}. Other
studies impose renewable orientation through sample selection and include no
firm-level portfolio measure
\citep{morina2021understanding,tsai2017factors}. Each of these measures is
defined using generated electricity rather than installed capacity. Samples
range from 37 producers in eight emerging European markets
\citep{tomczak2019} to 328 producers in six countries
\citep{DOPIERALA202211492}. Taken together, these differences leave no single
established empirical specification that can be transferred directly to our
setting.

To make the resulting specification choices transparent, we systematically
review the variables and empirical designs used in the related literature.
Appendix Table~\ref{tab:b1_studies} summarizes the outcomes, portfolio
measures, samples, and estimation approaches employed in previous studies.
We then compile the complete set of control variables used across these
studies and assess each variable according to the admissibility criteria
described in Appendix~\ref{app:covariates}. Appendix
Table~\ref{tab:b2_covariates} documents this screening process and records
whether, and for what reason, each variable is retained or excluded.

The composition of our sample guides the choice of the dependent variable.
Because the panel contains a substantial number of unlisted independent power
producers and state-owned firms, market-based outcomes such as Tobin's $Q$,
market capitalization, or the cost of capital cannot be constructed
consistently for all firms. We therefore focus on accounting-based measures of
profitability, which record realized rather than expected performance.\footnote{
The two families of measures capture distinct objects:
market-based indicators such as Tobin's $Q$ reflect investors'
forward-looking expectations about future cash flows, whereas accounting
measures record performance delivered over the preceding fiscal year
\citep{horvathova2010does}.}
Our primary outcome is return on assets, $\mathrm{ROA}_{ikt}$, which measures
accounting profitability relative to the firm's asset base and is less
mechanically affected by differences in leverage than equity-based measures.
As a complementary outcome, we use return on equity, $\mathrm{ROE}_{ikt}$,
which measures profitability relative to shareholder equity and is therefore
more sensitive to financing and capital-structure decisions.

Applying the screening procedure described in
Appendix~\ref{app:covariates} yields a parsimonious set of controls that can
be constructed consistently across listed, unlisted, and state-owned firms.
The baseline control vector, $X_{i,t-1}$, contains lagged firm size, leverage,
and sales growth, which account for central differences in firm scale,
financial structure, and recent operating performance. Entering these
variables with a one-year lag avoids conditioning on contemporaneous firm
outcomes and reduces the scope for reverse causation from current
profitability to the covariates.

ROA, ROE, leverage, and sales growth are winsorized symmetrically at the 1st
and 99th percentiles of the full panel. Firm size
enters in logarithms and is not winsorized. The bounded portfolio shares and
the constructed carbon-intensity measure are likewise not winsorized.
Table~\ref{tab:descriptives} provides definitions and summary statistics for
the outcomes, portfolio measures, and baseline controls.

Even after restricting attention to admissible covariates, uncertainty
remains over which combination of controls should enter the empirical model.
We therefore use Bayesian model averaging
\citep{hoeting1999bayesian,fernandez2001model} to assess sensitivity to alternative admissible model specifications and combination of control variables. Appendix~\ref{app:bma} describes the full
model-averaging procedure and reports the results.

\begin{table}[!tbp]
\centering
\caption{Sample descriptives.}
\label{tab:descriptives}
\begin{threeparttable}
\footnotesize
\begin{tabular}{lrrrrrr}
\toprule
Variable & Mean & SD & Min & Median & Max & $N$ \\
\midrule
ROA (\%, winsorized) & 4.04 & 9.38 & $-$40.88 & 2.82 & 51.22 & 11,971 \\
ROE (\%, winsorized) & 10.66 & 38.56 & $-$238.78 & 8.44 & 204.21 & 11,506 \\
Carbon intensity (tCO$_2$/MWh), $t-1$ & 0.150 & 0.243 & 0.000 & 0.000 & 1.010 & 11,971 \\
Gas capacity share (\%), $t-1$ & 20.61 & 36.81 & 0.00 & 0.00 & 100.00 & 11,971 \\
Carbon price (EUR/tCO$_2$) & 27.44 & 27.93 & 0.00 & 14.42 & 83.78 & 11,237 \\
Gas price (USD/MMBtu) & 11.16 & 8.57 & 2.67 & 8.86 & 37.52 & 11,971 \\
Firm size (log assets), $t-1$ & 12.06 & 2.62 & 2.32 & 11.86 & 20.46 & 11,971 \\
Sales growth (\%, winsorized), $t-1$ & 6.95 & 45.92 & $-$213.38 & 3.59 & 210.57 & 11,971 \\
Leverage (\%, winsorized), $t-1$ & 30.96 & 25.05 & 0.00 & 26.42 & 109.68 & 11,971 \\
Hydro capacity share (\%), $t-1$ & 19.36 & 37.27 & 0.00 & 0.00 & 100.00 & 11,971 \\
Nuclear capacity share (\%), $t-1$ & 1.60 & 10.14 & 0.00 & 0.00 & 100.00 & 11,971 \\
\bottomrule
\end{tabular}
\begin{tablenotes}[flushleft]
\footnotesize
\item \textit{Notes:} Summary statistics over the base sample of 11,971 firm-year observations on 1,357 firms, 2002--2023. 
Winsorized variables are capped at the 1st and 99th percentiles. Carbon intensity is defined in Equation~\eqref{eq:carbonintensity}. Capacity shares are firm-level shares of ownership-weighted installed MW, lagged one calendar year and bounded in $[0,100]$. ROE is defined only for firms with positive book equity, resulting in a slightly smaller sample; it enters the robustness analysis rather than the main specifications. The allowance price is reported over the years in which it is observed. The mechanism regressions begin with Phase~II in 2008.
\end{tablenotes}
\end{threeparttable}
\end{table}

\subsection{Classifying Technology Portfolios}
\label{subsec:classification}

European power firms operate highly heterogeneous generation portfolios,
ranging from portfolios concentrated in a single technology to portfolios
combining several generation sources. Comparing the profitability of firms
with different technology portfolios therefore requires translating
continuous technology shares into economically meaningful portfolio types.

The empirical literature has operationalized firms' technology orientation
in several ways. Some studies measure technology orientation through continuous renewable-generation shares or sort firms into discrete portfolio types using prespecified share cutoffs
\citep{DORIGONI2024114323,DOPIERALA202211492}, while others identify
portfolio patterns using unsupervised clustering
\citep{alova2020global}. Cutoff-based classifications are transparent and
easy to interpret (e.g., classifying a firm as renewable-oriented when
renewables account for at least 80 percent of its total electricity
generation), but results may depend on the selected threshold.
Unsupervised clustering is data-driven, but the obtained groups can be less straightforward to interpret.

We therefore implement a transparent rule-based classification as our baseline and an unsupervised clustering procedure as robustness check.
In the baseline, a firm is classified as predominantly renewable- or fossil-oriented based on the technology with the largest share of its annual generating capacity. 
Formally,
\begin{equation}
C_{it} = \operatorname*{arg\,max}_{m}\; s_{imt}.
\label{eq:classification}
\end{equation}
Firm-years in which the two largest technology shares are exactly equal are labelled as ties and not assigned to an archetype. There are three such firm-years in the estimation panel (0.03\%). We keep them in our analyses that do not use the archetype classification.

As a robustness check, we apply hierarchical clustering with Dynamic Time Warping to firms' portfolio-share trajectories. The resulting clusters agree with firms' modal dominant-technology class in 95\% of cases, with a nearly unchanged profitability gap. We therefore use the simple dominant-technology rule in the main analysis and report the clustering results in Appendix~\ref{app:cluster}.

We aggregate the dominant-technology classes into two portfolio archetypes. Firm-years dominated by wind or solar PV are classified as \textit{Renewable}, while firm-years dominated by gas, coal, or oil are classified as \textit{Fossil}. Firm-years dominated by hydro, nuclear, or biomass belong to neither archetype. They remain part of the broader panel but are excluded from regressions that directly compare Renewable- and Fossil-archetype firms.
Within this two-archetype sample, we define
\begin{equation}
\text{Renewable}_{it}
=
\begin{cases}
1, & \text{if } C_{it}\in\{\text{wind},\ \text{solar PV}\},\\
0, & \text{if } C_{it}\in\{\text{gas},\ \text{coal},\ \text{oil}\}.
\end{cases}
\label{eq:renewabledummy}
\end{equation}

The empirical specifications use the lagged indicator $\text{Renewable}_{i,t-1}$ to mitigate reverse causality concerns arising from contemporaneous portfolio adjustments (such as accelerated divestments of fossil or renewable assets following unexpected profit shocks), so that profitability in year $t$ is related to the portfolio held in the preceding year rather than to contemporaneous portfolio composition.

This discrete measure is used for the annual and regional profitability-gap analyses in Section~\ref{subsec:gap}. The mechanism analysis in Section~\ref{subsec:carbonprice} instead uses
portfolio carbon intensity and the gas capacity share.

\subsection{Estimating Profitability Gaps}
\label{subsec:gap}

To estimate how the conditional profitability difference between Renewable-
and Fossil-archetype firms evolves over time, we restrict the sample to these two archetypes and estimate
\begin{equation}
Y_{ikt}
=
\sum_{\tau\in\mathcal{T}}
\theta_{\tau}
\left(
\mathbbm{1}\{t=\tau\}
\times
\text{Renewable}_{i,t-1}
\right)
+
X_{i,t-1}'\gamma
+
\alpha_{kt}
+
\varepsilon_{ikt},
\label{eq:gap}
\end{equation}
where $Y_{ikt}$ denotes the profitability of firm $i$ in country $k$ and year $t$, measured as ROA in our baseline specification. Appendix~\ref{app:gap_robustness} shows robustness checks with ROE as the dependent variable.
$X_{i,t-1}$ is the lagged baseline control vector defined in Section~\ref{subsec:variables}, and $\alpha_{kt}$ is a country-by-year fixed effect. Standard errors are clustered by firm to allow arbitrary dependence across observations belonging to the same firm \citep{abadie2023}.

We include the interaction for every year in $\mathcal{T}$ and omit the time-invariant main effect of
$\text{Renewable}_{i,t-1}$, which is collinear with the full set of interactions. With fossil-archetype firms as the
reference category, each coefficient $\theta_{\tau}$
measures the conditional Renewable-minus-Fossil profitability difference in
year $\tau$ in levels, not relative to an omitted base year. 
A positive coefficient indicates higher profitability among Renewable-archetype firms.

The estimation window runs from 2002 to 2023. Constructing the lagged archetype requires the firm's portfolio classification in $t-1$, while lagged sales growth additionally requires revenue information for $t-2$. The first two years of the underlying panel therefore cannot contribute complete observations to this specification.

The country-by-year fixed effects absorb all observed and unobserved factors that are common to firms located in the same country and year. These include national energy policies, macroeconomic conditions, electricity demand, and country-level exposure to changes in the EU ETS. Each $\theta_{\tau}$ is consequently identified by comparing Renewable- and Fossil-archetype firms operating in the same country and year, conditional on their lagged size, leverage, and sales growth.

Generation portfolios are chosen rather than randomly assigned, and unobserved firm characteristics may influence both portfolio composition and profitability. 
The controls and country-by-year fixed effects mitigate but cannot fully eliminate this concern (Appendix~\ref{app:covariates} reports alternative control sets).
Thus, we interpret $\theta_{\tau}$ as a conditional association, not as the causal effect of converting a fossil portfolio into a renewable one.

We do not include firm fixed effects: because dominant portfolio types rarely change, they would absorb most of the variation in the archetype indicator, and identification therefore rests on differences between firms.
Section~\ref{subsec:carbonprice} complements this with a firm fixed-effects specification, in which within-firm variation comes from the interaction between continuous portfolio exposure and the time-varying allowance price.

To assess whether the profitability difference holds across Europe or is concentrated in particular regions, we assign each country to one of four EU sub-regions: Northern Europe (Sweden, Finland, Denmark, Estonia, Latvia, and Lithuania), Western Europe (Germany, France, Belgium, the Netherlands, Luxembourg, Austria, and Ireland), Southern Europe (Spain, Portugal, Italy, Greece, Croatia, Slovenia, Cyprus, and Malta), and Eastern Europe (Poland, Czechia, Hungary, Romania, Bulgaria, and Slovakia).

For each region $r$, we estimate a single-coefficient version of Equation~\eqref{eq:gap} over 2019--2023, the period in which a profitability gap emerges (Section~\ref{subsec:gap_results}):
\begin{equation}
Y_{ikt}
=
\beta^{r}\,
\text{Renewable}_{i,t-1}
+
X_{i,t-1}'\gamma
+
\alpha_{kt}
+
\varepsilon_{ikt},
\qquad
k(i)\in r,
\quad
t\in\{2019,\ldots,2023\}.
\label{eq:regional}
\end{equation}
Each regression uses only firms in region $r$. Thus, $\beta^{r}$ is the average conditional Renewable-minus-Fossil difference in that region over 2019--2023, allowing us to compare profitability gaps across regions. 
We estimate the same specification for the unrestricted EU-27 sample and report it as a pooled benchmark. Standard errors are clustered by firm within each regression.

\subsection{Estimating Carbon and Gas Price-Exposure Effects}
\label{subsec:carbonprice}

To examine whether the Renewable-Fossil profitability gaps are associated with firms' differential exposure to allowance and gas prices, we construct two measures from firms' installed generation portfolios. Portfolio carbon intensity provides a capacity-weighted benchmark of direct-combustion exposure. The gas-capacity share is defined as the firm's ownership-adjusted operating gas capacity divided by its total operating generation capacity. We interact portfolio carbon intensity with the allowance price and the gas-capacity share with the gas price.

For each fossil-fuel generating unit $u$ with a mapped fuel and generation
technology, we first express the fuel-specific combustion factor in tonnes of CO$_2$ per megawatt-hour of thermal input. The unit's benchmark
direct-combustion factor is then

\begin{equation}
e_u
=
\frac{
f_{\mathrm{fuel}(u)}
}{
\eta_{\mathrm{tech}(u)}
},
\label{eq:unit_emission_factor}
\end{equation}
where $f_{\mathrm{fuel}(u)}$ is the fuel-specific combustion factor in tonnes of CO$_2$ per megawatt-hour of thermal input and
$\eta_{\mathrm{tech}(u)}$ is the assumed net electrical efficiency of the
unit's generation technology. The resulting factor $e_u$ is measured in
tonnes of CO$_2$ per megawatt-hour of electricity.

The asset data record the fuel and generation technology of each unit. The
recorded fuel determines the applicable fuel-specific combustion factor,
while the generation technology determines the efficiency assumption. We
take the fuel factors from the IPCC default tables
\citep{ipcc2006guidelines} and convert them from tonnes of CO$_2$ per
terajoule to tonnes of CO$_2$ per megawatt-hour of thermal input, using
$1\ \mathrm{MWh}=0.0036\ \mathrm{TJ}$. The 2019 Refinement leaves the
stationary-combustion factors unchanged \citep{ipcc2019}, and the same
reference values appear in Annex~VI of the EU ETS Monitoring and Reporting
Regulation \citep{mrr2018}. For $\eta_{\mathrm{tech}(u)}$, we use the midpoint of the technology-specific net electrical efficiency range reported by \citet{jrc2012}. For technologies without a separately reported range, we use the analogues or family defaults. Appendix Table~\ref{tab:assumptions} reports the fuel-specific combustion factors, efficiency assumptions, resulting unit-level emission factors, and their sources.

We aggregate the unit-level factors to the firm-year level as

\begin{equation}
\mathrm{CI}_{i,t-1}
=
\frac{
\displaystyle
\sum_{u\in\mathcal{U}_{i,t-1}}
\mathrm{MW}_{iu,t-1}\,e_u
}{
\displaystyle
\sum_{u\in\mathcal{U}_{i,t-1}}
\mathrm{MW}_{iu,t-1}
},
\label{eq:carbonintensity}
\end{equation}

where $\mathcal{U}_{i,t-1}$ is the set of operating generation units held by
firm $i$ in year $t-1$, and $\mathrm{MW}_{iu,t-1}$ is unit $u$'s nameplate capacity multiplied by the firm's ownership share. Units that account for a larger share of the firm's operating capacity therefore receive greater weight.

Wind, solar PV, hydro, and nuclear units receive a direct-combustion factor
of zero. Biomass is also zero-rated in the baseline measure, following EU ETS accounting.\footnote{The zero rating follows the treatment of eligible
biomass under EU ETS accounting and does not imply that biomass has zero lifecycle emissions. Because the asset data do not identify whether each biomass unit satisfies the applicable sustainability criteria, the robustness analysis excludes biomass-dominated firms.} We use direct-combustion rather than lifecycle emissions because the EU ETS prices emissions released at regulated installations. Portfolio carbon intensity is a benchmark of the carbon exposure embedded in the firm's installed generation assets, not a measure of realized emissions or allowance obligations. It does not account for utilization, dispatch, plant-specific heat rates, fuel quality, or the allocation of emissions between electricity and heat in cogeneration plants. Similarly, the gas-capacity share measures the presence of gas-fired assets in the installed portfolio rather than realized gas generation, fuel consumption, or expenditure.

We estimate the association between portfolio exposure and the two prices
using
\begin{equation}
\begin{split}
Y_{ikt}
={}&
\beta_1
\left(
\mathrm{CI}_{i,t-1}
\times
\widetilde{\mathrm{CP}}_{t}
\right)
+
\beta_2
\left(
s^{\mathrm{gas}}_{i,t-1}
\times
\widetilde{\mathrm{GP}}_{t}
\right)
\\
&+
\beta_3 \mathrm{CI}_{i,t-1}
+
\beta_4 s^{\mathrm{gas}}_{i,t-1}
+
X_{i,t-1}'\gamma
+
\lambda_i
+
\delta_t
+
\varepsilon_{ikt},
\end{split}
\label{eq:mechanism}
\end{equation}
where $\widetilde{\mathrm{CP}}_{t}$ is the standardized annual EU ETS allowance price and $\widetilde{\mathrm{GP}}_{t}$ is the standardized annual European gas price. Both price series have mean zero and unit variance across firm-year observations in the common complete-case estimation sample; years are therefore weighted by the number of observations they contribute. The variable $s^{\mathrm{gas}}_{i,t-1}$ denotes the lagged gas-capacity share. The vector $X_{i,t-1}$ contains lagged firm size, leverage, sales growth, hydro share, and nuclear share. Firm fixed effects, $\lambda_i$, absorb time-invariant differences across firms, while year fixed effects, $\delta_t$, absorb shocks common to all firms in a given year.

The coefficient $\beta_1$ measures how the association between portfolio carbon intensity and profitability changes with the allowance price.
Because the allowance price is standardized, $\beta_1$ gives the differential change in ROA, in percentage points, associated with a one-standard-deviation increase in the allowance price for a one-tonne-per-MWh difference in portfolio carbon intensity. A negative coefficient implies that firms with more carbon-intensive portfolios perform worse relative to firms with less carbon-intensive portfolios as the allowance price rises.

The coefficient $\beta_2$ analogously measures how the association between the gas-capacity share and profitability changes with the gas price.
Since the gas-capacity share ranges from zero to one, $\beta_2$ corresponds to a unit difference in the gas-capacity share; the association for a ten-percentage-point difference is $0.1\beta_2$. Using the gas-capacity share for this interaction avoids assigning gas-price exposure to coal- and oil-fired assets that also contribute to portfolio carbon intensity.

We include both interactions jointly because allowance and gas prices rose
sharply during overlapping parts of the sample and firms with fossil-oriented portfolios may be exposed to both developments. Omitting either interaction could therefore attribute its association to the other price channel.

The estimation sample covers 2008--2023. The EU ETS began in 2005, but the
pilot phase from 2005 to 2007 is not covered by the consistent transaction
price series used in this analysis. The specification therefore begins with
Phase~II in 2008.

The interpretation of $\beta_1$ and $\beta_2$ requires that the interaction
terms do not proxy for omitted time-varying shocks that both co-move with the allowance or gas price and differentially affect firms according to their installed exposure. The controls and fixed effects reduce this concern but cannot eliminate it. We therefore interpret the coefficients as conditional price-exposure associations rather than as causal effects of allowance or gas prices.

Because exposure varies across firms and the two price series vary over time, standard errors are clustered at both the firm and year levels. The year dimension contains only sixteen clusters, for which conventional
cluster-robust inference may over-reject. We therefore additionally report
wild-cluster-bootstrap $p$-values based on year clusters
\citep{colin2015practitioner}.

Equation~\eqref{eq:mechanism} restricts the association between each portfolio exposure and profitability to vary linearly with the corresponding price. In year $t$, the implied coefficient on portfolio carbon intensity is $\beta_3+\beta_1\widetilde{\mathrm{CP}}_t$, and the implied coefficient on the gas-capacity share is
$\beta_4+\beta_2\widetilde{\mathrm{GP}}_t$. We examine whether these
associations instead vary more flexibly over time by replacing the
price-interaction terms and exposure main effects with separate exposure
coefficients for each year:

\begin{equation}
\begin{split}
Y_{ikt}
={}&
\sum_{\tau=2008}^{2023}
\theta^{\mathrm{CI}}_{\tau}
\left(
\mathbbm{1}\{t=\tau\}\times \mathrm{CI}_{i,t-1}
\right)
+
\sum_{\tau=2008}^{2023}
\theta^{\mathrm{gas}}_{\tau}
\left(
\mathbbm{1}\{t=\tau\}\times s^{\mathrm{gas}}_{i,t-1}
\right)
\\
&+
X_{i,t-1}'\gamma
+
\lambda_i
+
\delta_t
+
\varepsilon_{ikt}.
\end{split}
\label{eq:flexible_exposure}
\end{equation}

The sample, controls, and fixed effects match column~(3) of
Table~\ref{tab:carbon_intensity}. 
The standard errors are clustered at firm levels.

\section{Results}
\label{sec:results}

\subsection{Technology Portfolio Dynamics and Profitability Trends}

\begin{figure}[!t]
\centering
\includegraphics[width=0.75\textwidth]{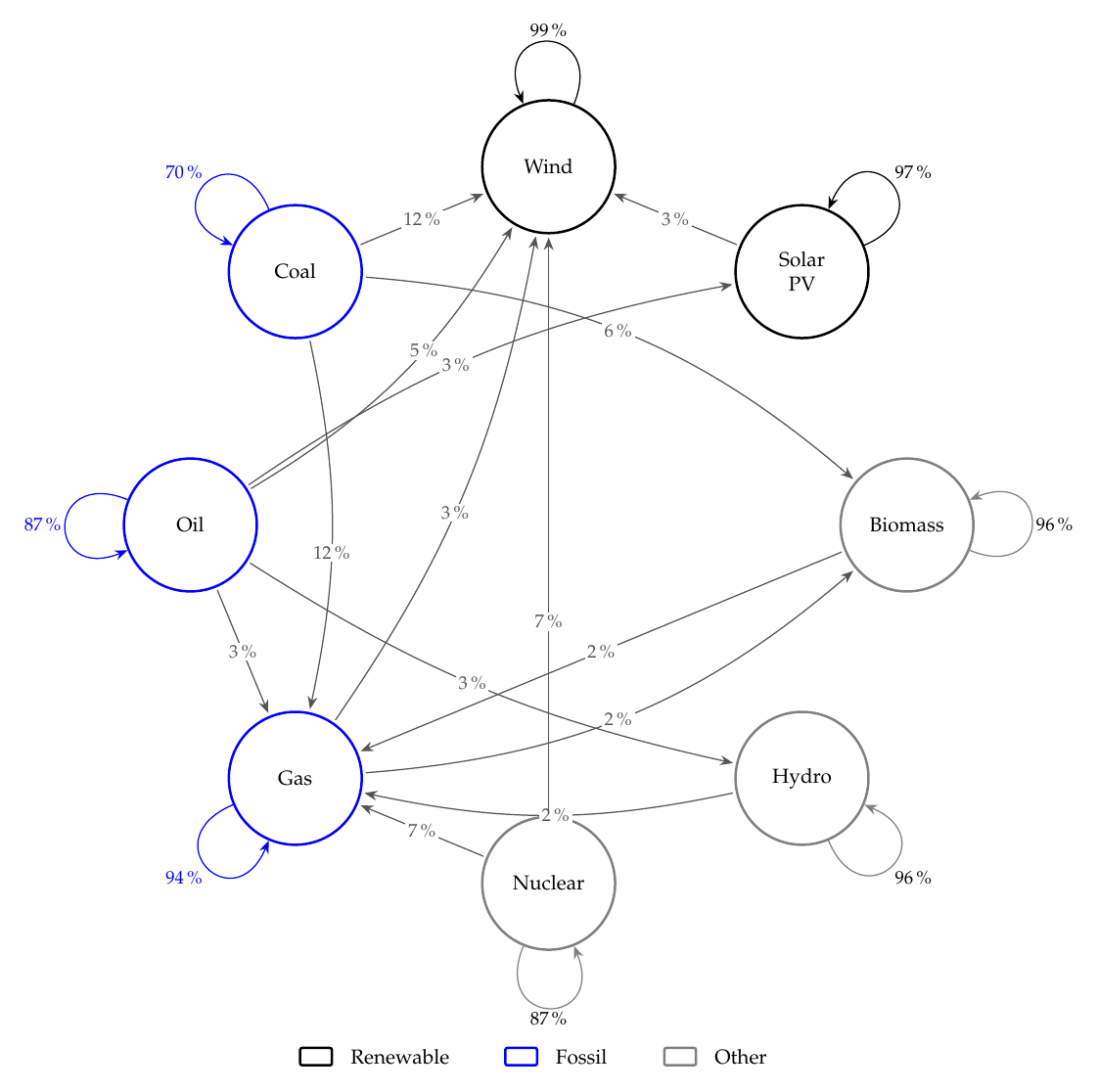}
\caption[Transition network]{%
{Transition network.}
Nodes are technology clusters and edges give the share of firms moving between clusters between the first and the last confirmed observation. Self-loops report the share retained and off-diagonal edges below 2\% are omitted. For the full transition matrix, see Appendix \ref{app:cluster} Figure \ref{fig:full_transition_matrix_classification}.%
}
\label{fig_1}
\end{figure}

We begin by examining how firms' generation portfolios evolve over time. In each year, we classify a firm by the technology with the largest share of its operating generation capacity. Figure~\ref{fig_1} visualizes the resulting transitions as a network, with nodes representing dominant technologies and edges showing the share of firms moving from their initial to their final dominant technology. 
The diagonal entries of the underlying transition matrix are shown as self-loops and give the share of firms that end in the same dominant technology in which they started. Thus, self-loops measure the persistence of the classification over a firm's observation window.
Dominant technology status is highly persistent. Self-transition probabilities range from 70\% for coal to 99\% for wind, with five of the eight technologies exceeding 94\%. Overall, approximately 96\% of the 1,107 firms with a confirmed classification at both ends of their observation window end in the technology in which they began.

Coal-dominant firms show the most movement across technologies.\footnote{A transition is defined as a change in a firm's single largest technology by capacity share. Note that this reflects a shift in portfolio dominance rather than a full replacement of the underlying generation fleet.} 
Among the 33 firms that begin as coal dominant, 4 transition to wind, 4 to gas, and 2 to biomass. By contrast, we observe no systematic transitions from other dominant-technology groups into coal. Gas-dominant portfolios are more persistent, with 94\% of firms retaining gas as their dominant technology. Gas is also the only fossil technology that receives a larger number of entrants from multiple other groups, including coal, oil, and nuclear, as well as hydro and biomass. Wind and solar portfolios are even more persistent, with 99\% and 97\% of firms, respectively, retaining their dominant technology. Moreover, wind is also the most common destination for firms transitioning out of other groups, with gas-dominant firms accounting for the largest source (6 firms). Overall, changes in dominant technology are rare, but the observed transitions are concentrated around shifts away from coal and towards gas and wind.

\begin{figure}[!t]
\centering
\includegraphics[width=\textwidth]{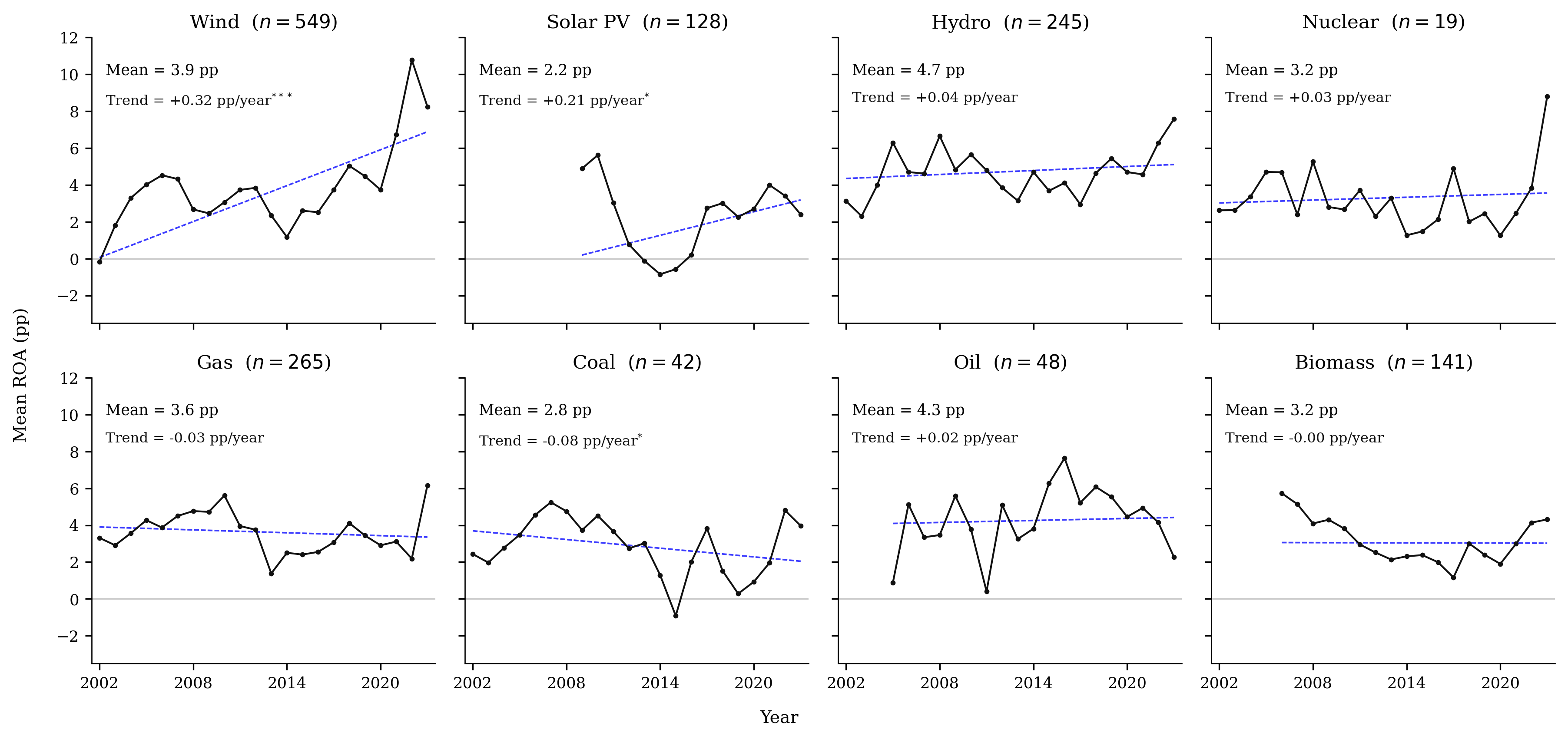}
\caption[Profitability levels and trends by dominant technology]{%
{Profitability levels and trends by dominant technology.}
Black lines report unweighted annual mean ROA by dominant-technology group and blue dashed lines show covariate-adjusted linear trends. Mean annotations average the displayed annual means, while trend annotations report
percentage-point changes per year. Panel titles report the number of distinct firms. Standard errors are clustered by firm.
$^{*}p<0.10$, $^{**}p<0.05$, $^{***}p<0.01$.%
}
\label{fig_2}
\end{figure}

Figure~\ref{fig_2} relates the different generation portfolios to firm profitability, by reporting the unweighted annual mean ROA for each dominant-technology group. 
Average profitability differs across technologies, with mean ROA ranging from 2.2\% for solar PV to 4.7\% for hydropower. Oil- and wind-dominated firms also have relatively high mean ROA, at 4.3\% and 3.9\%, respectively, while coal-dominated firms average 2.8\%. The remaining technologies fall between these values. 
Thus, average profitability levels reveal no systematic divide between renewable- and fossil-dominated firms.\footnote{Excluding the 2022 energy crisis leaves the broad pattern unchanged: mean ROA ranges from 4.6\% for hydropower firms to 2.1\% for solar firms.}

Beyond these differences in profitability levels, technologies also exhibit
distinct long-run dynamics. The blue dashed lines in Figure~\ref{fig_2}
estimate these trends on all firm-year observations while controlling for
lagged sales growth, firm size, leverage, and country fixed effects. Wind
firms display the strongest upward trajectory, with ROA increasing by 0.32
percentage points per year ($p<0.01$). This result remains positive and
statistically significant in the pooled and firm-fixed-effects specifications reported in Appendix Table~\ref{tab:technology_trends}. 
Coal firms exhibit a smaller but consistently negative trend.
The estimated solar trend is positive in the baseline specification but less
precisely estimated across alternative specifications. Hydro, nuclear, gas,
oil, and biomass exhibit no consistently detectable linear trends.

Taken together, this analysis delivers three main findings. First, dominant
technology orientations are highly persistent, with portfolio shifts
concentrated among a small number of firms. Second, profitability levels
differ meaningfully across technology groups, although these differences do
not follow a simple renewable-fossil ordering. Third, the technologies follow distinct long-run trajectories: wind exhibits a robust positive trend, coal a modest negative trend, while gas and oil remain statistically flat. The solar trajectory declines initially and recovers in later years, but does not exhibit a robust linear trend.

\subsection{A Profitability Gap Emerges After 2018}
\label{subsec:gap_results}

\begin{figure}[!t]
\centering
\includegraphics[width=\textwidth]{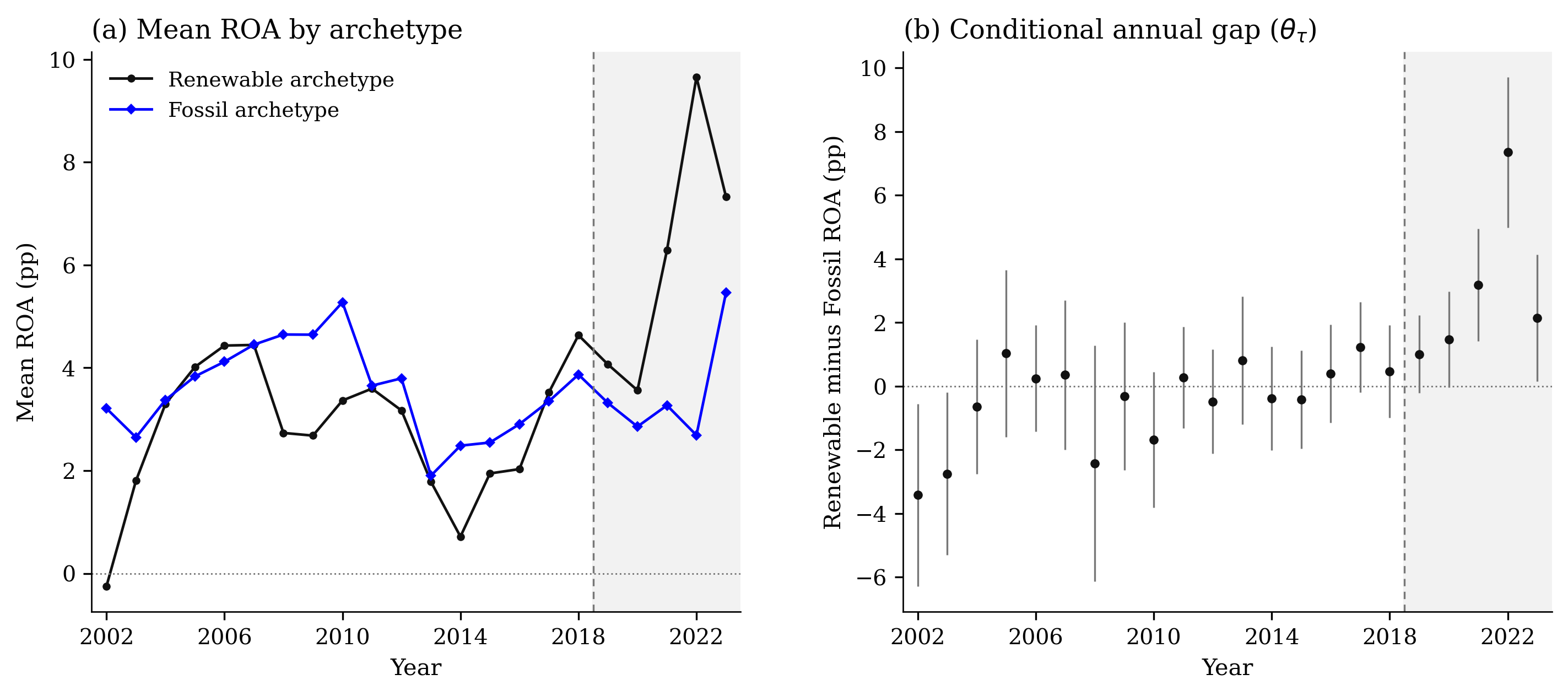}
\caption[The widening profitability gap]{%
{The widening profitability gap.}
Panel A reports unweighted annual mean ROA for Renewable- and Fossil-archetype firms. Panel B reports the annual conditional profitability gap from Equation~\ref{eq:gap}, with ROA as the dependent variable. Error bars are 95\% confidence intervals based on firm-clustered standard errors. The shaded area marks 2019--2023, and the dashed line marks the end of 2018.%
}
\label{fig_3}
\end{figure}

We next show that a Renewable-Fossil profitability gap emerges after 2018.
Panel~A of Figure~\ref{fig_3} reports the unweighted annual mean return on
assets for Renewable- and Fossil-archetype firms. Fossil-archetype firms have higher mean ROA in most years through 2016, with the main exceptions occurring in 2005 and 2006. Their profitability advantage narrows after 2010, and the ranking reverses in 2017. Renewable-archetype firms subsequently have higher mean ROA in every remaining year, with the largest differences occurring in 2021 and 2022.

To assess whether this descriptive pattern persists after conditioning on differences in firm characteristics and shocks common to firms operating in
the same national market, Panel~B of Figure~\ref{fig_3} reports the annual
conditional profitability gap estimated from Equation~\eqref{eq:gap}. Each
coefficient compares Renewable- and Fossil-archetype firms operating in the
same country and year, conditional on lagged firm size, leverage, and sales
growth. Estimates are negative in 2002 and 2003, and then fluctuate around zero from 2004 through 2018 without evidence of a systematic profitability gap during that period. From 2019 onward, the estimated gap turns positive, increases through 2022, and remains positive in 2023.

The annual coefficients describe how the profitability gap evolves over time, but do not directly test whether its average level changes after 2018.
Panel~A of Table~\ref{tab:gap_prepost} therefore reports a pooled pre/post specification that formally compares the Renewable-Fossil gap before and after this date. 
Over the pre-2019 period, the estimated gap is effectively
zero, at $-0.032$ percentage points ($SE=0.418$; $p=0.939$). After 2018, the gap increases by 3.227 percentage points ($SE=0.625$; $p<0.001$), implying an average Renewable-Fossil difference of 3.195 percentage points over
2019--2023 ($SE=0.572$; $p<0.001$). The interaction estimate therefore shows that the positive post-2018 gap represents a statistically significant change relative to the preceding period.

The estimated widening is robust to alternative break years, fixed-effects
structures, clustering levels, and sample definitions. In particular, it
remains positive and statistically significant when we separately exclude
parent companies, subsidiaries, and firms owning only a single generating
unit. Across all specifications using ROA, the estimated post-break increase
ranges from 2.896 to 3.632 percentage points and remains statistically
significant at the 1 percent level. The result is also positive and
statistically significant when profitability is measured using ROE
($p=0.036$). 
Appendix Tables~\ref{tab:gap_rob_overview}--\ref{tab:gap_rob_single} report the full results.

\begin{table}[!t]
\centering
\caption{The Renewable-Fossil profitability gap.}
\label{tab:gap_prepost}
\begin{threeparttable}
\footnotesize
\begin{tabular}{@{}lrrrr@{}}
\toprule
\multicolumn{5}{@{}l}{\textit{Panel A: Gap before and after 2019}} \\
\addlinespace
 & Estimate & Standard error & Firms & Firm-years \\
\midrule
Gap before 2019    & $-$0.032 & 0.418 & 989 & 8,359 \\
Change after 2018  & 3.227*** & 0.625 &  &  \\
Gap 2019--2023     & 3.195*** & 0.572 &  &  \\
\addlinespace
\addlinespace
\multicolumn{5}{@{}l}{\textit{Panel B: Gap by region, 2019--2023}} \\
\addlinespace
 & Estimate & Standard error & Firms & Firm-years \\
\midrule
EU-27 pooled       & 3.261*** & 0.592 & 844 & 3,107 \\
Northern Europe    & 2.184* & 1.247 & 134 & 466 \\
Western Europe     & $-$0.018 & 0.484 & 370 & 1,353 \\
Southern Europe    & 7.501*** & 1.218 & 307 & 1,165 \\
Eastern Europe     & 6.128** & 2.511 & 33 & 123 \\
\midrule
Country $\times$ year FE & Yes & & & \\
Lagged firm controls     & Yes & & & \\
\bottomrule
\end{tabular}

\begin{tablenotes}[flushleft]\footnotesize
\item \textit{Notes:}
ROA and all estimates are in percentage points. Panel A reports the pre-2019
gap, the post-2018 change, and the implied 2019--2023 gap from a single
regression. Panel B reports separate estimates of
Equation~\eqref{eq:regional} for the EU-27 and each region over 2019--2023.
Standard errors are clustered by firm and stars are based on these standard
errors. $^{*}p<0.10$, $^{**}p<0.05$, $^{***}p<0.01$.
\end{tablenotes}
\end{threeparttable}
\end{table}

We next ask whether the post-2018 profitability gap between Renewable- and
Fossil-archetype firms is concentrated in particular European regions or
extends more broadly across the EU-27. Panel~B of
Table~\ref{tab:gap_prepost} reports the regional estimates alongside the
pooled EU-27 result. For the EU-27 as a whole, Renewable-archetype firms have 3.261 percentage points higher ROA than Fossil-archetype firms
($SE=0.592$; $p<0.001$), closely matching the post-period estimate from the
pre/post specification.

The largest regional gap is found in Southern Europe, where
Renewable-archetype firms have 7.501 percentage points higher ROA than
Fossil-archetype firms ($SE=1.218$; $p<0.001$). The point estimates are also
positive in Eastern and Northern Europe. The Eastern European estimate is
6.128 percentage points ($SE=2.511$; $p=0.020$), although it is based on only 123 firm-years and 33 firms. The Northern European gap is 2.184 percentage points but is less precisely estimated ($SE=1.247$; $p=0.082$). Western Europe is the exception: its estimate is close to zero, at $-0.018$
percentage points ($SE=0.484$; $p=0.970$). The regional evidence is therefore strongest in Southern Europe, with additional positive evidence in Eastern Europe that should be interpreted in light of its small sample.

\subsection{The Role of Carbon and Gas Price Exposures}
\label{subsec:carbonprice_results}

We next assess whether, and to what extent, the widening
profitability gap documented in the previous section is
associated with firms' differential exposure to two recent major developments in
European energy markets: the post-2018 rise in EU ETS allowance prices and
the gas-price shock associated with the 2021--2022 European energy crisis. 
Firms differ substantially in the carbon intensity and gas exposure of their generation
portfolios, so these common price movements may be associated with different
profitability outcomes across firms.

\begin{figure}[!t]
\centering
\includegraphics[width=\textwidth]{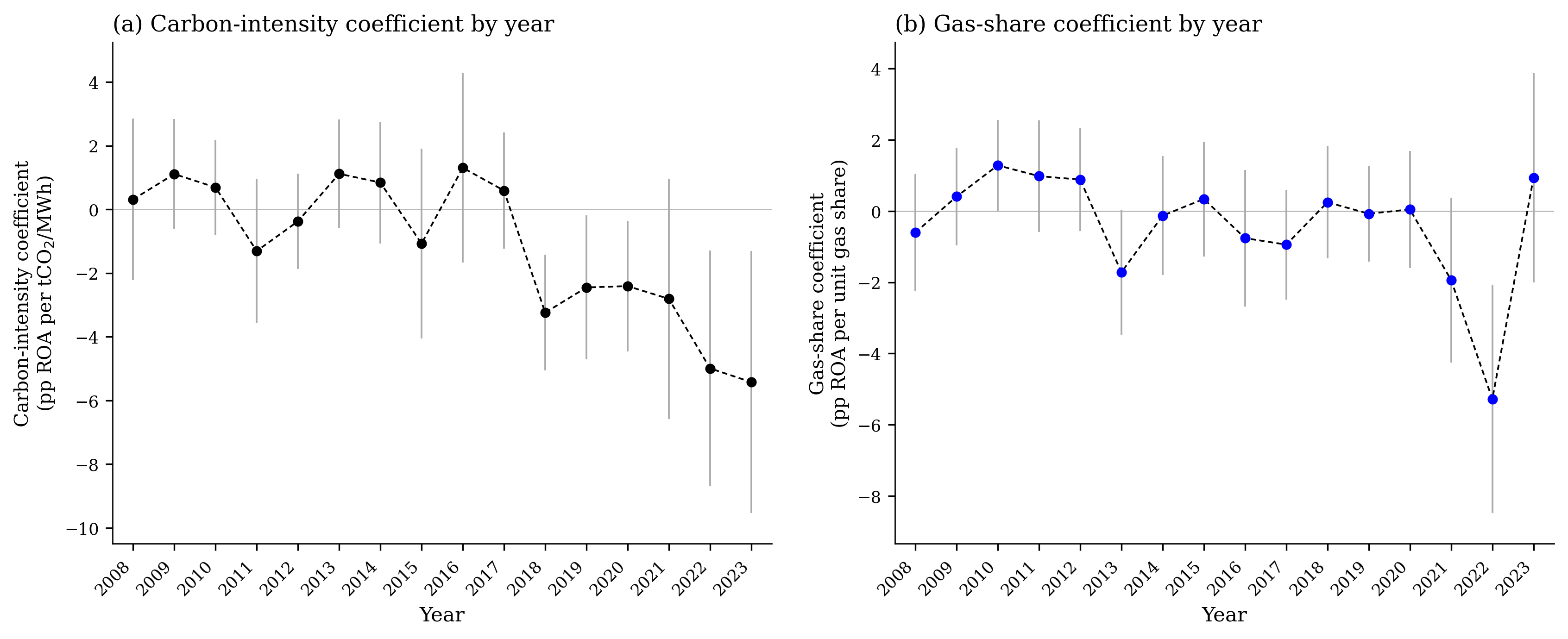}
\caption[Time-varying carbon-intensity and gas-exposure coefficients]{%
{Time-varying carbon-intensity and gas-exposure coefficients.}
Panels~(a) and~(b) report the jointly estimated annual carbon-intensity and gas-exposure coefficients from Equation~\eqref{eq:flexible_exposure}, after subtracting each series' 2008--2018 mean. Markers denote point estimates, dashed lines connect adjacent annual estimates, and error bars are 95\% confidence intervals clustered by firm. %
}
\label{fig:flexible_coefficients}
\end{figure}

Figure~\ref{fig:flexible_coefficients} reports the annual carbon-intensity and gas-exposure coefficients estimated from Equation~\eqref{eq:flexible_exposure}. The two coefficient series are
estimated jointly and expressed relative to their respective means over
2008--2018. Each coefficient therefore measures how the profitability
associated with a given exposure in that year differs from its average association over 2008--2018. 

Panel~(a) displays the annual carbon-intensity coefficients. Through 2017, estimates fluctuate around their pre-period mean without a clear trend. The coefficient turns negative in 2018 and remains persistently below zero thereafter, becoming most pronounced toward the end of the sample. In 2022 and 2023, a 0.1 tCO$_2$/MWh increase in portfolio carbon intensity is associated with approximately 0.50 and 0.54 percentage points lower ROA, respectively, relative to the 2008--2018 benchmark. A joint Wald test rejects the null that the five post-2018 deviations are jointly zero ($\chi^2(5)=16.87$; $p=0.0047$).

In contrast, Panel (b) reveals a short-lived gas-exposure effect concentrated in the 2022 price crisis. The annual coefficients display little systematic deterioration through 2020, become
more negative in 2021, and fall sharply in 2022, when European gas prices
reach their sample maximum. In 2023, the coefficient returns close to its
pre-period mean as gas prices decline. These post-2018 deviations are again jointly significant ($\chi^2(5)=13.82$; $p=0.0168$), but unlike the persistent carbon-intensity coefficient, the adverse gas-exposure effect is confined to the 2022 price crisis.
Thus, we find that unlike the persistent late-sample carbon-intensity coefficient, the adverse gas-exposure coefficient is concentrated in the 
European gas-price crisis.

\begin{table}[!t]
\centering
\caption{Profitability, carbon intensity and the allowance price.}
\label{tab:carbon_intensity}
\begin{threeparttable}
\footnotesize
\begin{tabular}{@{}lccc@{}}
\toprule
 & Gas & Carbon & \textbf{Both} \\
 & only & only & \textbf{jointly} \\
 & (1) & (2) & \textbf{(3)} \\
\midrule
Carbon intensity $\times$ CP & --- & $-$2.680*** & $-$1.756*** \\
 &  & (0.833) & (0.457) \\
 & --- & [0.0002] & [0.0026] \\
\addlinespace
Gas share $\times$ GP & $-$1.785*** & --- & $-$1.371*** \\
 & (0.286) &  & (0.296) \\
\midrule
Firm fixed effects        & Yes & Yes & Yes \\
Year fixed effects        & Yes & Yes & Yes \\
Lagged baseline controls  & Yes & Yes & Yes \\
Hydro and nuclear shares  & Yes & Yes & Yes \\
Within $R^2$              & 0.0253 & 0.0238 & 0.0271 \\
Firms                     & 1,265 & 1,265 & 1,265 \\
Firm-years                & 10,888 & 10,888 & 10,888 \\
Year clusters             & 16 & 16 & 16 \\
\bottomrule
\end{tabular}
\begin{tablenotes}[flushleft]\footnotesize
\item \textit{Notes:}
The dependent variable is ROA in percentage points. Carbon intensity is defined in Equation~\eqref{eq:carbonintensity}. Both prices are standardized over the estimation sample. Parentheses report standard errors clustered by firm and year. Brackets report year-cluster wild-bootstrap $p$-values (9,999 replications) for the carbon interaction. 
*** $p<0.01$, ** $p<0.05$, * $p<0.10$.
\end{tablenotes}
\end{threeparttable}
\end{table}

We next examine whether the carbon intensity and gas shares are associated with allowance and gas prices, respectively. Table~\ref{tab:carbon_intensity} reports the resulting full 2008--2023 sample estimates.
Column~(2) first estimates the carbon-price relationship without including the gas-price interaction. The coefficient on Carbon Intensity $\times$ Carbon Price is $-2.680$, with a wild-cluster-bootstrap $p$-value of 0.0002.
Column~(3) includes both price-exposure interactions. The carbon-price coefficient declines in absolute value to $-1.756$ but remains statistically significant under wild-cluster-bootstrap inference ($p=0.0026$). Because the
allowance price is standardized, the joint estimate implies that, for two portfolios differing in carbon intensity by 0.1 tCO$_2$/MWh, a one-standard-deviation higher allowance price is associated with ROA being 0.176 percentage points lower for the more carbon-intensive portfolio relative to the less carbon-intensive portfolio.

The gas-price coefficient is $-1.785$ when estimated separately in column~(1) and declines in absolute value to $-1.371$ when the carbon-price interaction is included in column~(3). The joint estimate implies that, for
two portfolios differing in their gas-capacity share by ten percentage points, a one-standard-deviation higher gas price is associated with ROA being 0.137 percentage points lower for the more gas-exposed portfolio relative to the less gas-exposed portfolio. Both gas coefficients are statistically significant at the 1 percent level based on standard errors
clustered by firm and year.\footnote{A wild-cluster-bootstrap $p$-value is not reported for the gas interaction because its restricted bootstrap
distribution is not centred.}

Carbon intensity and gas-capacity share are measured in different units, so their coefficients are not directly comparable. We therefore express both in percentage points of the Renewable-Fossil profitability gap, multiplying each coefficient by (i) the exposure difference between Fossil- and Renewable-archetype firms (0.479 tCO$_2$/MWh in carbon intensity; 68.2 percentage points in gas-capacity share) and (ii) the change in the corresponding average price between 2008--2018 and 2019--2023.

In 2019--2023, the gap averages 3.26 percentage points above its 2008--2018 mean. The carbon-price component accounts for 1.29 percentage points (about 40\%) of this widening, and the gas-price component for 0.61 percentage points (about 19\%). At the point estimates, the two terms jointly account for 1.90 percentage points of the widening, or about 60\%; the remaining 1.36 percentage points are not captured by these price-exposure terms.

The direction of these effects is robust to alternative exposure measures. Replacing carbon intensity with firms' Renewable and Fossil capacity shares yields the same pattern: higher carbon prices are associated with lower relative profitability among fossil-heavy firms and higher relative profitability among renewable-heavy firms (Appendix Table~\ref{tab:alt_exposure}).

Taken together, we find that the two price channels differ in both size and timing. The carbon-price channel is larger and persistent, operating across the entire post-2018 period. The gas-price channel is about half as large and transient, confined to the 2022 crisis. Jointly, they account for roughly 60 percent of the average widening.

\section{Conclusion}
\label{sec:conclusion}

Generation portfolios are a persistent source of heterogeneity across power firms: only about 4\% of firms in our panel change their dominant technology over two decades. We show that this inertia shapes how changes in the market environment reach firms' accounts. Renewable- and fossil-oriented firms earned similar returns until 2018. Thereafter, a gap of about 3.2 percentage points in return on assets opened, concentrated in Southern Europe and absent in Western Europe.

The timing of this divergence helps interpret the mixed evidence in the existing literature. Estimates that pool over the whole period impose a stable relationship between renewable orientation and performance and can therefore obscure the central empirical pattern. 
Renewable-oriented firms were not consistently more profitable over the past two decades. Their relative profitability advantage emerges only in the later years.

Rising allowance prices are the clearest correlate of the divergence. Carbon-intensive portfolios lost relative profitability as allowance prices rose, and this pattern persists through 2023 and alone accounts for about 40 percent of the profitability widening. Gas exposure adds roughly another 20 percent, but this association is confined to the 2022 crisis and not robust to wild-cluster bootstrap inference. The remaining part of the widening, together with the regional differences, points to factors our price terms do not capture, such as market design, support schemes, or resource quality.

Several limitations should be considered when interpreting our results. Our measure of carbon intensity is based on installed capacity rather than realized generation or emissions. Moreover, all estimates are conditional associations rather than causal effects. The profitability differences therefore reflect how inherited portfolios were exposed to changing market conditions, not the returns to firms adjusting their portfolios.

Our results nonetheless add an important dimension to the previous findings that the EU ETS has had little effect on the average regulated firm. Within the power sector, the same carbon-price change can have markedly different effects on profitability depending on firms’ inherited asset portfolios. Whether these differences accelerate the transition depends on how firms respond: higher relative returns may strengthen both the incentive and the internal capacity to invest in renewables, yet the portfolio transitions we document have so far been slow. How retained earnings, ownership, and regulation shape this reallocation is a natural question for future work.

\newpage

\section*{Data Availability}

The firm-level financial and power-plant data used in this study were
obtained from S\&P Capital IQ and the S\&P Global Power Plant Database
under licence and cannot be redistributed by the authors. The carbon-
and gas-price series are obtained from the sources described in the
paper. Information on data construction and access is provided in the
replication package.

\section*{Code Availability}
The code will be made public upon publication.


\section*{Funding}
RF was supported by the Austrian Federal Ministry for Innovation, Mobility and Infrastructure (BMIMI) under the endowed professorship for "Data-Driven Knowledge Generation: Climate Action".  
AP acknowledges support by the OeNB anniversary fund under contract number 18943 and the NSF APTO project under contract number 2403999. 

\section*{Competing Interests}

The authors declare no competing interests.


\bibliographystyle{elsarticle-harv}
\bibliography{references}

\clearpage
\appendix
\setcounter{page}{1}
\counterwithin{table}{section}
\counterwithin{figure}{section}
\counterwithin{equation}{section}
\renewcommand{\thetable}{\Alph{section}\arabic{table}}
\renewcommand{\thefigure}{\Alph{section}\arabic{figure}}
\renewcommand{\theequation}{\Alph{section}.\arabic{equation}}

\thispagestyle{plain}

\begin{center}
    {\Large\bfseries Online Appendix for}\\[1.2em]
    {\Large\bfseries ``The Evolution of Technology Portfolios and Firm Profitability \\ in the European Power Sector''}\\[2.5em]
    {\large Robin Fischer and Anton Pichler}\\[2.5em]
    {\large \monthyeardate\today}
\end{center}
\vspace{3em}

\renewcommand{\contentsname}{Contents}
\setcounter{tocdepth}{2}
{\hypersetup{linkcolor=black}
\startcontents[appendix]
\printcontents[appendix]{}{1}{}
}

\clearpage
\section{Data and Sample Construction}
\label{app:data}

\subsection{Data Sources and Ownership Matching}
\label{app:data_sources}

This appendix provides additional information on the construction of the
firm-level panel. We combine financial accounts from S\&P Capital IQ with
unit-level generation data from the S\&P Global Power Plant Database. The two sources share a firm identifier, which allows us to assign generation units to the legal entities recorded as their direct owners. Nameplate capacity is multiplied by the firm's ownership stake before being aggregated to the firm-year level.

Following \citet{alova2020global}, we conduct the analysis at the level of the direct asset owner rather than aggregating generation assets to the ultimate parent company. The literature provides no clear consensus on the extent to which subsidiaries' investment and operating decisions are determined by their parent companies rather than by local market, regulatory, and business conditions. Subsidiaries may operate in different countries or electricity subsectors from their parents and may therefore face different incentives when investing in, retaining, or retiring generation assets.

Aggregation to the ultimate-parent level could also obscure economically
meaningful differences in generation portfolios. A diversified parent may
own subsidiaries with distinct technology portfolios, so combining their
assets would mechanically smooth differences in renewable and fossil
orientation. We therefore argue that the direct-owner level provides a more
consistent match between the generation assets attributed to an entity and
the financial accounts used to measure its profitability, while preserving
the subsidiary-level portfolio heterogeneity central to our analysis.

\newpage
\subsection{Variable Construction}
\label{app:variables}

\begin{table}[!htbp]
\centering
\caption{Variables from S\&P Global used for the analysis}
\label{tab:a1_variables}
\begin{threeparttable}
\footnotesize
\renewcommand{\arraystretch}{0.96}
\begin{tabular}{@{}llp{7.7cm}@{}}
\toprule
S\&P field & Notation & Description \\
\midrule
\multicolumn{3}{@{}l}{\textit{Panel A: Market Intelligence power plant data (generating unit)}} \\
\addlinespace
\texttt{OWNER\_KEY}                & $i$              & Owning company identifier; matches \texttt{SP\_ENTITY\_ID} in the financial pull \\
\texttt{UNIT\_KEY}                 & $u$              & Generating unit identifier \\
\texttt{UNIT\_NAMEPLATE\_CAPACITY} &                  & Nameplate capacity of the unit, MW \\
\texttt{OWN}                       &                  & Ownership stake of the owner in the unit, per cent \\
\texttt{PRIMARY\_FUEL\_TYPE}       & fuel$(u)$        & Primary fuel burned; mapped to eight technology families \\
\texttt{GEN\_TECH}                 & tech$(u)$        & Generation technology (steam turbine, gas turbine, combined cycle) \\
\texttt{COUNTRY\_CODE}             &                  & Country of the unit; EU-27 retained \\
\texttt{OWN\_STATUS}               &                  & Unit status; \textit{Current} and \textit{Retired} retained \\
\texttt{YR\_UNIT\_IN\_SVC}         &                  & Year the unit entered service \\
\texttt{YR\_UNIT\_RETIRED\_FROM\_SVC} &               & Year the unit left service, missing while operating \\
\addlinespace
\addlinespace
\multicolumn{3}{@{}l}{\textit{Panel B: Capital IQ company financials (firm-year)}} \\
\addlinespace
\texttt{SP\_ENTITY\_ID}            & $i$              & Firm identification number \\
\texttt{IQ\_NET\_INC\_PAR}      &                     & Net income   \\
\texttt{IQ\_TOTAL\_ASSETS}         &                  & Total assets \\
\texttt{IQ\_TOTAL\_REV}            &                  & Total revenue \\
\texttt{IQ\_TOTAL\_DEBT}           &                  & Total debt \\
\texttt{IQ\_TOTAL\_EQUITY}         &                  & Total equity \\
\addlinespace
\addlinespace
\multicolumn{3}{@{}l}{\textit{Panel C: Variables constructed from the above}} \\
\addlinespace
                                   & $Y_{ikt}$        & Return on assets: \texttt{IQ\_NET\_INC\_PARENT} over \texttt{IQ\_TOTAL\_ASSETS}, winsorised at the 1st and 99th percentiles \\
                                   & $\text{MW}_{u}$  & Ownership-weighted unit capacity, \texttt{UNIT\_NAMEPLATE\_CAPACITY} $\times$ \texttt{OWN}$/100$ \\
                                   & $e_u$            & Unit emission factor from fuel and technology; see Eq.~\eqref{eq:unit_emission_factor} \\
                                   & $\text{CI}_{i,t-1}$ & Capacity-weighted carbon intensity of the firm's portfolio; see Eq.~\eqref{eq:carbonintensity} \\
                                   & $\text{GasShare}_{i,t-1}$ & Gas capacity over total capacity, both ownership-weighted \\
                                   & $C_{it}$         & Dominant technology cluster: the family holding the largest share of $\sum_u \text{MW}_u$ \\
                                   & $X_{i,t-1}$      & Baseline controls: log total assets, total debt over total assets, and the annual change in log total revenue. The price-exposure models additionally include hydro and nuclear capacity shares \\
\bottomrule
\end{tabular}

\begin{tablenotes}[flushleft]\footnotesize
\item \textit{Notes:}
Panel A is the unit-level power plant extract and Panel B the Capital IQ financial
pull; the two are linked by \texttt{OWNER\_KEY} $=$ \texttt{SP\_ENTITY\_ID}. 
\end{tablenotes}
\end{threeparttable}
\end{table}

\newpage 

\subsection{Sample Construction}
\label{app:sample}

\begin{table}[!htbp]
\centering
\caption{From the Raw Extracts to the Estimation Samples}
\label{tab:funnel}
\begin{threeparttable}
\footnotesize
\begin{tabular}{@{}lrr@{}}
\toprule
 & Observations & Firms \\
\midrule
\multicolumn{3}{@{}l}{\textit{Panel A: Generating units}} \\
\addlinespace
Raw unit records in the extract & 281,759 & 36,757 \\
In an EU-27 country, owner and capacity recorded & 73,055 & 10,111 \\
Fuel type mapped to one of eight technology families & 71,394 & 9,905 \\
\addlinespace
\addlinespace
\multicolumn{3}{@{}l}{\textit{Panel B: Firm-year capacity panel and merge with financials}} \\
\addlinespace
Unit-years between commissioning and retirement & 1,176,351 &  \\
Aggregated to firm-year capacity & 173,783 & 8,596 \\
Matched to a Capital IQ financial record & 66,743 & 5,136 \\
ROA finite and total assets positive & 54,686 & 4,379 \\
At least three annual observations per firm & 54,297 & 4,107 \\
Years through 2023 & 51,469 & 4,107 \\
Complete on all regression variables & 28,872 & 3,123 \\
\addlinespace
\addlinespace
\multicolumn{3}{@{}l}{\textit{Panel C: Estimation samples}} \\
\addlinespace
Base sample after industry restriction and cluster assignment & 11,971 & 1,357 \\
Renewable- or fossil-dominant in $t-1$ & 8,484 & 989 \\
Gap regressions (Table~\ref{tab:gap_prepost}) & 8,359 & 986 \\
Mechanism regressions (Table~\ref{tab:carbon_intensity}) & 10,888 & 1,265 \\
\bottomrule
\end{tabular}

\begin{tablenotes}[flushleft]\footnotesize
\item \textit{Notes:}
Rows count generating units in Panel A and firm-years from the second row of
Panel B onward; the unit-year row is the intermediate expansion of each unit over
the years it was in service and has no firm count. Within the EU-27 every record
already carried a valid owner, a positive capacity and a status of
\textit{Current} or \textit{Retired}, so those screens are folded into one line.
Panel C is not a single chain: the mechanism
sample is larger than the gap sample because it does not require a firm to be
renewable- or fossil-dominant in $t-1$. The gap regressions lose a further 125
rows that sit in country-year cells containing a single observation, which a
country-by-year fixed effect cannot use.
\end{tablenotes}
\end{threeparttable}
\end{table}

\newpage

\subsection{Country Sample}
\begin{table}[!htbp]
\centering
\caption{Sample Composition by Country and European Subregion}
\label{tab:country_counts}
\begin{threeparttable}
\footnotesize
\begin{tabular}{@{}lrrr@{}}
\toprule
 & Firms & Firm-years & Share of firm-years \\
\midrule
\textit{Northern Europe} & & & \\
\quad Sweden & 124 & 1,001 & 8.4\% \\
\quad Finland & 88 & 396 & 3.3\% \\
\quad Denmark & 32 & 159 & 1.3\% \\
\quad Estonia & 4 & 46 & 0.4\% \\
\quad Lithuania & 3 & 47 & 0.4\% \\
\quad Latvia & 1 & 20 & 0.2\% \\
\cmidrule(l){2-4}
\quad \textit{Subtotal} & 252 & 1,669 & 13.9\% \\
\addlinespace
\textit{Western Europe} & & & \\
\quad Germany & 338 & 3,521 & 29.4\% \\
\quad France & 139 & 1,417 & 11.8\% \\
\quad Belgium & 30 & 141 & 1.2\% \\
\quad Austria & 29 & 349 & 2.9\% \\
\quad Ireland & 20 & 147 & 1.2\% \\
\quad Luxembourg & 5 & 80 & 0.7\% \\
\quad Netherlands & 5 & 49 & 0.4\% \\
\cmidrule(l){2-4}
\quad \textit{Subtotal} & 566 & 5,704 & 47.6\% \\
\addlinespace
\textit{Southern Europe} & & & \\
\quad Spain & 249 & 1,884 & 15.7\% \\
\quad Italy & 173 & 1,586 & 13.2\% \\
\quad Portugal & 61 & 554 & 4.6\% \\
\quad Slovenia & 5 & 69 & 0.6\% \\
\quad Greece & 3 & 31 & 0.3\% \\
\quad Cyprus & 1 & 15 & 0.1\% \\
\quad Malta & 1 & 19 & 0.2\% \\
\cmidrule(l){2-4}
\quad \textit{Subtotal} & 493 & 4,158 & 34.7\% \\
\addlinespace
\textit{Eastern Europe} & & & \\
\quad Poland & 12 & 110 & 0.9\% \\
\quad Romania & 12 & 94 & 0.8\% \\
\quad Bulgaria & 7 & 62 & 0.5\% \\
\quad Hungary & 7 & 78 & 0.7\% \\
\quad Czechia & 5 & 61 & 0.5\% \\
\quad Slovakia & 3 & 35 & 0.3\% \\
\cmidrule(l){2-4}
\quad \textit{Subtotal} & 46 & 440 & 3.7\% \\
\midrule
Total & 1,357 & 11,971 & 100.0\% \\
\bottomrule
\end{tabular}

\begin{tablenotes}[flushleft]\footnotesize
\item \textit{Notes:}
Counts refer to the base sample of Table~\ref{tab:funnel}, Panel C. 
\end{tablenotes}
\end{threeparttable}
\end{table}

\newpage

\subsection{Portfolio Carbon Intensity}
\label{app:carbon_intensity}

Table~\ref{tab:assumptions} reports the fuel-side emission factors,
technology-specific efficiency assumptions, and resulting unit-level
combustion factors used to construct portfolio carbon intensity.

\begin{table}[H]
\centering
\caption{Inputs to the unit-level emission factor}
\label{tab:assumptions}
\begin{threeparttable}
\footnotesize
\begin{tabular}{@{}llrrlrr@{}}
\toprule
 & & EU-27 & $f$ & $\eta$ range & $\eta$ & $e$ \\
Fuel & Technology & GW & (tCO$_2$/TJ) & (source) & used & (tCO$_2$/MWh) \\
\midrule
\multicolumn{7}{@{}l}{\textit{Panel A: coal}} \\
\addlinespace
Lignite            & Steam turbine & 28.1 & 101.0 & 32--40\% (a) & 0.360 & 1.010 \\
Anthracite         & Steam turbine & 22.0 &  98.3 & 32--40\% (a) & 0.360 & 0.983 \\
Coal, unspecified  & Steam turbine & 80.1 & 101.0 & 32--40\% (a) & 0.360 & 1.010 \\
Bituminous coal    & Steam turbine &  3.4 &  94.6 & 32--40\% (a) & 0.360 & 0.946 \\
Sub-bituminous     & Steam turbine &  0.7 &  96.1 & 32--40\% (a) & 0.360 & 0.961 \\
\addlinespace
\multicolumn{7}{@{}l}{\textit{Panel B: natural gas}} \\
\addlinespace
Natural gas & Combined cycle       & 146.3 & 56.1 & 52--60\% (b) & 0.560 & 0.361 \\
Natural gas & Open-cycle turbine   &  27.3 & 56.1 & 35--42\% (c) & 0.385 & 0.525 \\
Natural gas & Steam turbine        &  19.2 & 56.1 & ---   (d)    & 0.360 & 0.561 \\
\addlinespace
\multicolumn{7}{@{}l}{\textit{Panel C: oil}} \\
\addlinespace
Fuel oil & Steam turbine    & 18.5 & 77.4 & ---   (d)    & 0.360 & 0.774 \\
Fuel oil & Gas turbine      &  9.8 & 77.4 & 30--40\% (e) & 0.350 & 0.796 \\
Fuel oil & Combined cycle   &  1.4 & 77.4 & ---   (d)    & 0.500 & 0.557 \\
\addlinespace
\multicolumn{7}{@{}l}{\textit{Panel D: zero-rated}} \\
\addlinespace
\multicolumn{2}{@{}l}{Wind, solar, hydro, nuclear, biomass} & --- & 0 & --- & --- & 0 \\
\bottomrule
\end{tabular}
\begin{tablenotes}[flushleft]\footnotesize
\item \textit{Notes:}

\item \textit{Fuel factors.} IPCC 2006 Guidelines, Volume~2 (Energy), Chapter~2 (Stationary
Combustion), Table~2.2 \citep{ipcc2006guidelines}. The 2019 Refinement records ``no refinement'' for
stationary-combustion factors \citep{ipcc2019}, and the same values appear in Annex~VI of
Commission Implementing Regulation (EU) 2018/2066 \citep{mrr2018}, the schedule against which
the firms in this panel surrender allowances.

\item \textit{Efficiency sources}, all \citet{jrc2012}:
(a) p.~11, the range reported for the installed fleet;
(b) pp.~3 and 19;
(c) pp.~3 and 14;
(d) not tabulated separately --- held to the closest sourced analogue, see below;
(e) p.~14, liquid-fuel gas turbines.
We take the median of each reported range.

\item \textit{Why the fleet range for coal.} The same report gives two figures. Across
technology vintages it reports 33--47\% (p.~3), median 40\%. For the installed fleet it states
that ``the majority of pulverised coal plants are more than 15 years old and operate with
sub-critical steam parameters and efficiencies between 32--40\%'' (p.~11), median 36\%. This
panel covers 2002--2023, so the fleet figure is the relevant one; the technology-range median
would implicitly assume the fleet is spread uniformly across vintages, which the same source
denies.

\item \textit{The three cells marked} (d). Gas- and oil-fired steam turbines run the same
Rankine cycle as the coal steam turbine and take its 36\%; oil combined cycle takes 50\%. These
are the authors' assumptions by analogy, not values from the cited source. Together they cover
39.1~GW. Internal-combustion units (7.5~GW gas, 4.8~GW oil, under 3\% of fossil capacity) have
no reported range and fall through to their fuel family's capacity-weighted average.

\item \textit{The unspecified coal label.} 80.1~GW, 59\% of EU-27 coal capacity, carries no
fuel sort. The baseline treats it as lignite; the alternative treats it as other-bituminous.
Re-estimating the mechanism on each, the carbon component moves by 0.007 percentage points,
or 0.3 percentage points of the accounted share. Sensitivities that hold the coefficient fixed
overstate this by roughly a factor of eight, because raising a factor scales the intensity
variable up and the coefficient proportionally down.
\end{tablenotes}
\end{threeparttable}
\end{table}

\section{Variable Selection and Model Uncertainty}
\label{app:extended_results}

\subsection{Covariate Selection}
\label{app:covariates}

We construct the candidate set of control variables from prior studies of
renewable investment and power-firm profitability. Table~\ref{tab:b1_studies}
summarizes the studies included in this review. Table~\ref{tab:b2_covariates}
records each distinct covariate and explains whether it enters the candidate
set, is absorbed by the fixed effects, is excluded by design, or cannot be
constructed consistently for the full sample.

\begin{table}[htbp]
\centering
\caption{Studies relating renewable investment to power-firm profitability}
\label{tab:b1_studies}
\begin{threeparttable}
\footnotesize
\setlength{\tabcolsep}{4pt}
\begin{tabularx}{\textwidth}{@{}l R{1.00} c l R{0.72} R{1.28}@{}}
\toprule
Study & Sample & Period & Source & Method & Dependent variables \\
\midrule
\citet{RUGGIERO20173676} & 66 listed utilities, 26 countries & 2005--14 & DataStream & FE, RE & ROE, ROA, Tobin's~$q$ \\
\addlinespace
\citet{tsai2017factors} & 93 firms, 34 countries & 2008--13 & COMPUSTAT & Pooled panel & ROA, gross margin, interest coverage \\
\addlinespace
\citet{tomczak2019} & 37 firms, 8 European emerging markets & 2008--17 & EMIS & Cluster analysis & None; 16 ratios compared \\
\addlinespace
\citet{morina2021understanding} & 41 listed firms, EU & 2004--18 & Eikon & OLS, RE, GMM & ROA, Tobin's~$q$ \\
\addlinespace
\citet{DOPIERALA202211492} & 328 firms, 6 countries & 2011--19 & Orbis & FE, RE & ROA, ROE \\
\addlinespace
\citet{DORIGONI2024114323} & 64 listed utilities, continental Europe & 2011--20 & Refinitiv & OLS & ROA, WACC, Tobin's~$q$ \\
\midrule
\textit{This paper} & 1,357 firms, EU-27 & 2002--23 & Capital IQ & FE, BMA & ROA, ROE \\
\bottomrule
\end{tabularx}
\begin{tablenotes}[flushleft]\footnotesize
\item \textit{Notes:} FE and RE denote fixed- and random-effects panel estimators.
\citet{tsai2017factors} describe their estimator only as panel data methodology and do not
specify the panel structure further. Variable definitions and estimating equations were
read from each paper directly rather than from secondary summaries.
\end{tablenotes}
\end{threeparttable}
\end{table}

\begin{table}[H]
\centering
\caption{Literature-based covariate universe and treatment in this paper}
\label{tab:b2_covariates}
\begin{threeparttable}
\footnotesize
\begin{tabular}{@{}lll@{}}
\toprule
Covariate & Used in & Treatment or reason \\
\midrule
\addlinespace
\multicolumn{3}{@{}l}{\textit{Included in the candidate set} (14)} \\
\quad Firm size (log total assets) & DP22 &  \\
\quad Leverage (debt / assets) & RL17, M21, DP22 &  \\
\quad Debt-to-equity & D24 &  \\
\quad Sales growth & RL17, M21, D24 &  \\
\quad Asset turnover (revenue / assets) & DP22, TT17 & TT17 enters the reciprocal \\
\quad Renewable capacity share & D24 & DP22 enters a 50 percent indicator \\
\quad Fossil capacity share & DP22 & entered as a continuous share \\
\quad Hydro capacity share & DP22 & entered as a continuous share \\
\quad Legal form / listing status & DP22 &  \\
\quad National renewable share in generation & RL17, TT17 &  \\
\quad Industrial electricity price & M21 &  \\
\quad National energy consumption & TT17 &  \\
\quad GDP growth & M21 &  \\
\quad Inflation & M21 &  \\
\addlinespace
\multicolumn{3}{@{}l}{\textit{Added by this paper} (3)} \\
\quad Nuclear capacity share & -- & completes the portfolio partition \\
\quad EU ETS carbon price & -- & the mechanism this paper studies \\
\quad Natural gas price & -- & the leading alternative explanation \\
\addlinespace
\multicolumn{3}{@{}l}{\textit{Absorbed by the fixed effects} (3)} \\
\quad Country indicators & DP22 & Country-by-year fixed effects \\
\quad Linear time trend & RL17 & Year fixed effects \\
\quad Crisis-period indicators & M21 & Year fixed effects \\
\addlinespace
\multicolumn{3}{@{}l}{\textit{Excluded from the common-sample model space} (5)} \\
\quad EBIT & D24 & Component of the outcome \\
\quad Lagged profitability & M21 & Dynamic specification \\
\quad Renewable generation volume & RL17 & Alternative parameterisation \\
\quad Road-network length & TT17 & No plausible channel \\
\quad Firm age & M21 & Available for only 55 percent of the panel \\
\addlinespace
\multicolumn{3}{@{}l}{\textit{Unavailable at the firm level} (10)} \\
\quad Firm size (market capitalisation) & RL17, M21, TT17 & Requires a listing \\
\quad Beta & D24 & Requires a listing \\
\quad CO2 scope-1 emissions & D24 & Voluntary disclosure \\
\quad Carbon intensity (emissions / assets) & RL17 & Voluntary disclosure \\
\quad Energy-efficiency policy indicator & D24 & Voluntary disclosure \\
\quad Current ratio & D24 & Not reported at direct-owner level \\
\quad Depreciation over assets & DP22 & Not reported at direct-owner level \\
\quad Capital intensity (capex / sales) & RL17, M21 & Not reported at direct-owner level \\
\quad R\&D expenditure & TT17 & Not reported at direct-owner level \\
\quad Employees & D24, TT17 & Snapshot only, not a panel \\
\addlinespace
\multicolumn{3}{@{}l}{\textit{Unavailable as a consistent EU-27 panel} (4)} \\
\quad Wholesale power price & DP22 & Absorbed by country-by-year FE \\
\quad Market concentration & M21 & Absorbed by country-by-year FE \\
\quad Support schemes (feed-in tariff, certificates, quota) & RL17, M21 & Absorbed by country-by-year FE \\
\quad Tax incentives & TT17 & Absorbed by country-by-year FE \\
\addlinespace
\multicolumn{3}{@{}l}{\textit{Not applicable to an EU-27 sample} (1)} \\
\quad Developed-market indicator & RL17 & Sample is entirely EU-27 \\
\bottomrule
\end{tabular}
\begin{tablenotes}[flushleft]\footnotesize
\item \textit{Notes:} Study keys: D24 \citet{DORIGONI2024114323}, DP22 \citet{DOPIERALA202211492},
M21 \citet{morina2021understanding}, RL17 \citet{RUGGIERO20173676}, TT17 \citet{tsai2017factors}. A construct is
counted once where several studies use it, and variables appearing only as dependent
variables are excluded.
\item The five regression studies use 37 distinct regressors:
14 enter the candidate set,
3 are absorbed by the fixed effects,
5 are excluded from the common-sample model space,
10 cannot be measured at the firm
level, 4 are unavailable as a
consistent country panel, and 1 does
not apply to an EU-27 sample. We additionally consider 3 variables
motivated by the institutional setting. The admissibility criteria are stated in the text.
\end{tablenotes}
\end{threeparttable}
\end{table}

\subsection{Bayesian Model Averaging}
\label{app:bma}

We use Bayesian model averaging \citep{hoeting1999bayesian,fernandez2001model}
to assess uncertainty over the admissible control set. The analysis uses one complete-case sample for all seventeen candidates and compares three model spaces: no fixed effects, year fixed
effects, and country-by-year fixed effects. Within each structure, every
candidate not absorbed by its fixed effects can enter or leave the models. We
use Zellner's $g$-prior at the BRIC benchmark \citep{fernandez2001benchmark},
a beta-binomial model prior \citep{ley2009effect}, and exact enumeration of the model space. Table~\ref{tab:bma} reports the posterior inclusion probabilities. 

\begin{table}[H]
\centering
\caption{Variable definitions and data sources for the BMA}
\label{tab:a2_variables}
\begin{threeparttable}
\footnotesize
\begin{tabular}{@{}lp{5.2cm}ccl@{}}
\toprule
Variable & Definition & Timing & Units & Source \\
\midrule
\multicolumn{5}{@{}l}{\textit{Outcomes}} \\
\quad Return on assets & Net income attributable to parent over total assets & $t$ & \% & Capital IQ \\
\addlinespace
\multicolumn{5}{@{}l}{\textit{Generation portfolio}} \\
\quad Renewable capacity share & Wind plus solar MW over total MW & $t-1$ & 0--1 & S\&P plant data \\
\quad Fossil capacity share & Coal, gas and oil MW over total MW & $t-1$ & 0--1 & S\&P plant data \\
\quad Hydro capacity share & Hydro MW over total MW & $t-1$ & 0--1 & S\&P plant data \\
\quad Nuclear capacity share & Nuclear MW over total MW & $t-1$ & 0--1 & S\&P plant data \\
\addlinespace
\multicolumn{5}{@{}l}{\textit{Firm controls}} \\
\quad Firm size & Natural log of total assets & $t-1$ & log & Capital IQ \\
\quad Leverage & Total debt over total assets & $t-1$ & 0--1 & Capital IQ \\
\quad Sales growth & $\log(\text{revenue}_{t-1}) - \log(\text{revenue}_{t-2})$ & $t-1$ & log points & Capital IQ \\
\quad Debt-to-equity & Total debt over total equity, equity $>0$ & $t-1$ & ratio & Capital IQ \\
\quad Asset turnover & Total revenue over total assets & $t-1$ & ratio & Capital IQ \\
\quad Listing status & Indicator: company type is a public company & fixed & 0/1 & Capital IQ \\
\addlinespace
\multicolumn{5}{@{}l}{\textit{Prices and country conditions}} \\
\quad EU ETS carbon price & Calendar-year mean of daily EUA settlement prices & $t$ & EUR/tCO$_2$ & ICE \\
\quad Natural gas price & Calendar-year mean of the European spot index & $t$ & USD/MMBtu & FRED \\
\quad Renewable share in generation & Renewable share of gross electricity consumption & $t$ & \% & Eurostat \\
\quad Industrial electricity price & Band ID (2,000--19,999 MWh), excluding taxes and levies, calendar-year mean & $t$ & EUR/kWh & Eurostat \\
\quad Final energy consumption & Final energy consumption, Energy Efficiency Directive definition, all sectors & $t$ & log ktoe & Eurostat \\
\quad GDP growth & Annual growth of real GDP & $t$ & \% & World Bank \\
\quad Inflation & Annual growth of the consumer price index & $t$ & \% & World Bank \\
\bottomrule
\end{tabular}
\begin{tablenotes}[flushleft]\footnotesize
\item \textit{Notes:}
Country-level series are retrieved from the following datasets: renewable share in
generation, Eurostat \texttt{nrg\_ind\_ren} (\texttt{REN\_ELC}); industrial electricity
price, Eurostat \texttt{nrg\_pc\_205} for 2007 onward and \texttt{nrg\_pc\_205\_h} before;
final energy consumption, Eurostat \texttt{nrg\_bal\_s} (\texttt{FEC\_EED}); GDP growth,
World Bank \texttt{NY.GDP.MKTP.KD.ZG}; inflation, World Bank \texttt{FP.CPI.TOTL.ZG}.
\end{tablenotes}
\end{threeparttable}
\end{table}

\begin{table}[!htbp]
\centering
\caption{Posterior Inclusion Probabilities from Bayesian Model Averaging}
\label{tab:bma}
\begin{threeparttable}
\footnotesize
\begin{tabular}{@{}llrrr@{}}
\toprule
 & & No fixed & Year & Country $\times$ year \\
Covariate & Source & effects & FE & FE \\
\midrule
Firm size (log assets) & Dopierala 2022 / Tomczak 2019 & 1.000 & 1.000 & 1.000 \\
Leverage (debt/assets) & Ruggiero 2017 / Morina 2021 & 1.000 & 1.000 & 1.000 \\
EU ETS carbon price & own & 1.000 & absorbed & absorbed \\
Debt-to-equity & Dorigoni 2024 & 1.000 & 1.000 & 1.000 \\
Renewable share in generation & Ruggiero 2017 / Tsai 2017 & 0.999 & 0.940 & absorbed \\
Sales growth & Ruggiero 2017 / Morina 2021 & 0.997 & 0.977 & 0.969 \\
Renewable capacity share & Dorigoni 2024 & 0.891 & 0.866 & 0.664 \\
Hydro capacity share & Dopierala 2022 & 0.848 & 0.831 & 0.679 \\
GDP growth & Morina 2021 & 0.820 & 1.000 & absorbed \\
Asset turnover & Dopierala 2022 / Tomczak 2019 & 0.277 & 0.325 & 0.113 \\
Nuclear capacity share & novel & 0.232 & 0.231 & 0.196 \\
Fossil capacity share & Dopierala 2022 & 0.121 & 0.149 & 0.352 \\
Industrial electricity price & Morina 2021 & 0.063 & 0.205 & absorbed \\
Natural gas price & own & 0.034 & absorbed & absorbed \\
Publicly listed company & Dopierala 2022 & 0.015 & 0.021 & 0.018 \\
Final energy consumption & Tsai 2017 & 0.014 & 0.017 & absorbed \\
Inflation (CPI) & Morina 2021 & 0.014 & 0.026 & absorbed \\
\bottomrule
\end{tabular}

\begin{tablenotes}[flushleft]\footnotesize
\item \textit{Notes:}
Bayesian model averaging of ROA over the covariates used in the panel studies of
this literature, with Zellner's $g$-prior at the BRIC benchmark, a beta-binomial
model prior, and the model space enumerated exactly. All three structures are
estimated on one shared sample, so the only difference across columns is what the
fixed effects absorb. A covariate varying only at an absorbed level is marked
\textit{absorbed}: it cannot be identified in that column, which is not the same as
being uninformative. The source column gives the study each candidate is taken
from.
\end{tablenotes}
\end{threeparttable}
\end{table}

\newpage

\subsection{Portfolio Classification}
\label{app:cluster}

This subsection provides additional evidence on the classification of firms'
generation portfolios. It reports the cluster solution, the technology
composition of each cluster, and the agreement between the cluster-based and
dominant-technology classifications.

\begin{table}[!htbp]
\centering
\caption{Dominant-Technology Rule against DTW Clusters}
\label{tab:dtw_crosstab}
\begin{threeparttable}
\footnotesize
\begin{tabular}{@{}lrrrrrrrr@{}}
\toprule
 & Wind & Solar PV & Biomass & Hydro & Nuclear & Gas & Oil & Coal \\
\midrule
Wind & 519 & 0 & 1 & 3 & 0 & 3 & 2 & 3 \\
Solar PV & 7 & 112 & 0 & 1 & 0 & 0 & 1 & 0 \\
Biomass & 1 & 0 & 124 & 4 & 0 & 0 & 0 & 1 \\
Hydro & 0 & 0 & 0 & 223 & 3 & 6 & 0 & 1 \\
Nuclear & 0 & 0 & 0 & 0 & 16 & 0 & 0 & 2 \\
Gas & 0 & 0 & 4 & 3 & 0 & 222 & 0 & 17 \\
Oil & 0 & 0 & 0 & 0 & 0 & 5 & 38 & 0 \\
Coal & 0 & 0 & 1 & 0 & 0 & 0 & 0 & 34 \\
\bottomrule
\end{tabular}

\begin{tablenotes}[flushleft]\footnotesize
\item \textit{Notes:} Rows give each firm's modal annual class under the
dominant-technology rule, columns its DTW cluster. Off-diagonal cells are firms the two
procedures classify differently; the diagonal holds 95 per cent of firms.
\end{tablenotes}
\end{threeparttable}
\end{table}

\begin{table}[!htbp]
\centering
\caption{Headline Estimates under Two Portfolio Classifications}
\label{tab:dtw_robustness}
\begin{threeparttable}
\footnotesize
\begin{tabular}{@{}lcc@{}}
\toprule
 & Dominant rule & DTW clusters \\
 & (1) & (2) \\
\midrule
Change after 2018 & 3.227 & 3.113 \\
 & (0.625) & (0.624) \\
Gap 2019--2023 & 3.195 & 3.180 \\
 & (0.572) & (0.573) \\
Carbon intensity $\times$ CP & $-$1.756 & $-$1.756 \\
Exposure gap $\Delta CI$ & 0.4787 & 0.4719 \\
Carbon-price component & 1.294 & 1.276 \\
\midrule
Firms & 986 & 971 \\
Firm-years & 8,359 & 8,309 \\
\bottomrule
\end{tabular}

\begin{tablenotes}[flushleft]\footnotesize
\item \textit{Notes:}
Column (1) classifies each firm-year by its largest technology share, as in the main
text. Column (2) replaces that rule with dynamic time warping on the portfolio-share
trajectories followed by average-linkage hierarchical clustering at $K=8$, with the
archetype taken from the cluster's dominant technology. The two agree for 95 per
cent of firms. The coefficient on carbon intensity is identical across columns because
the mechanism equation does not use the classification; what the classification changes
is the exposure gap $\Delta CI$ that converts it into a component of the profitability
gap. Standard errors clustered by firm in parentheses.
\end{tablenotes}
\end{threeparttable}
\end{table}

\begin{figure}[H]
\centering
\includegraphics[width=0.65\textwidth]{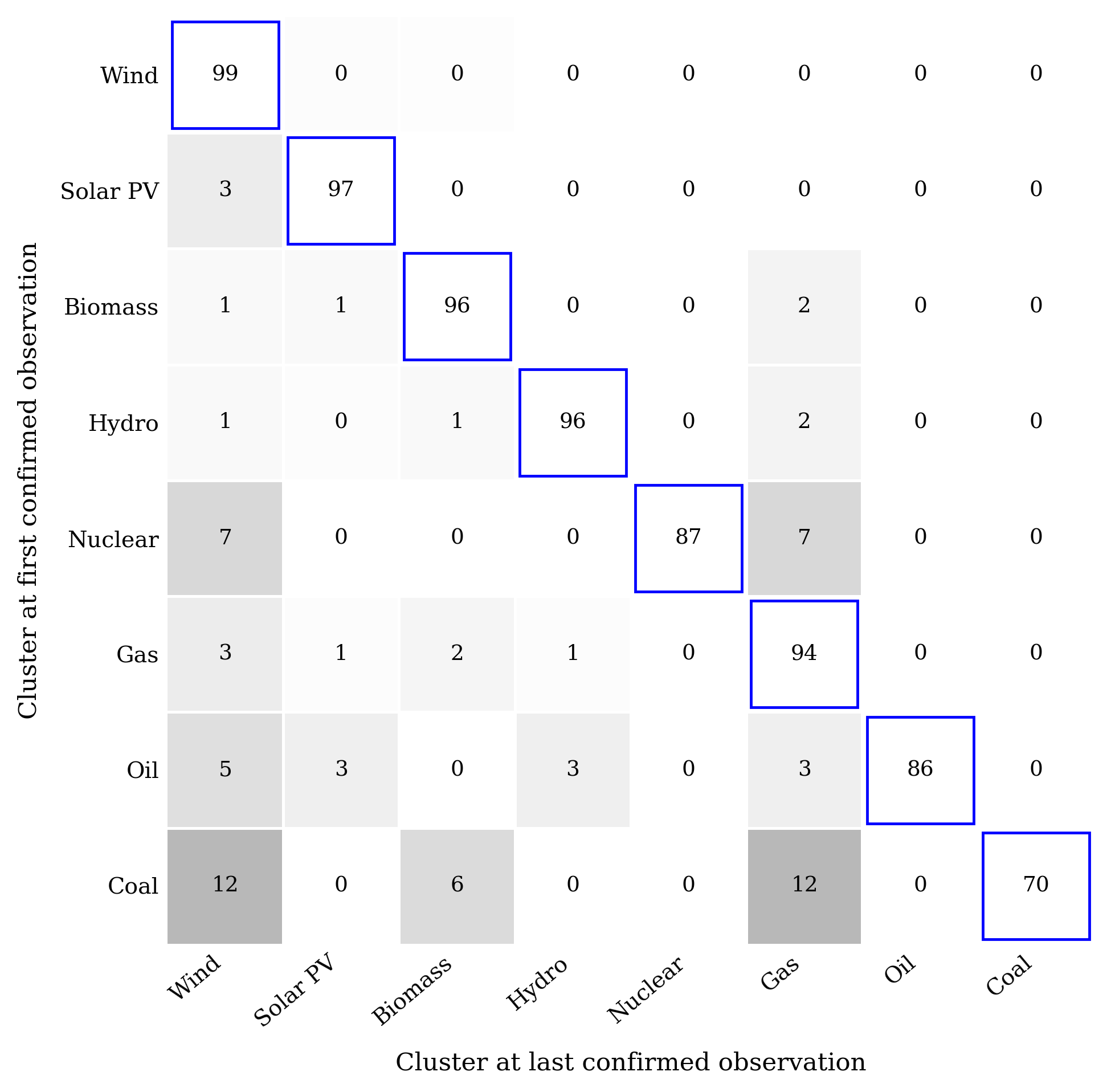}
\caption[Full Transition Matrix of dominant technology classification]{%
{Full Transition Matrix of dominant technology classification.}
Rows sum to 100 per cent; displayed values are rounded to integers and may not add up.%
}
\label{fig:full_transition_matrix_classification}

\end{figure}

\begin{figure}[!t]
\centering
\includegraphics[width=0.65\textwidth]{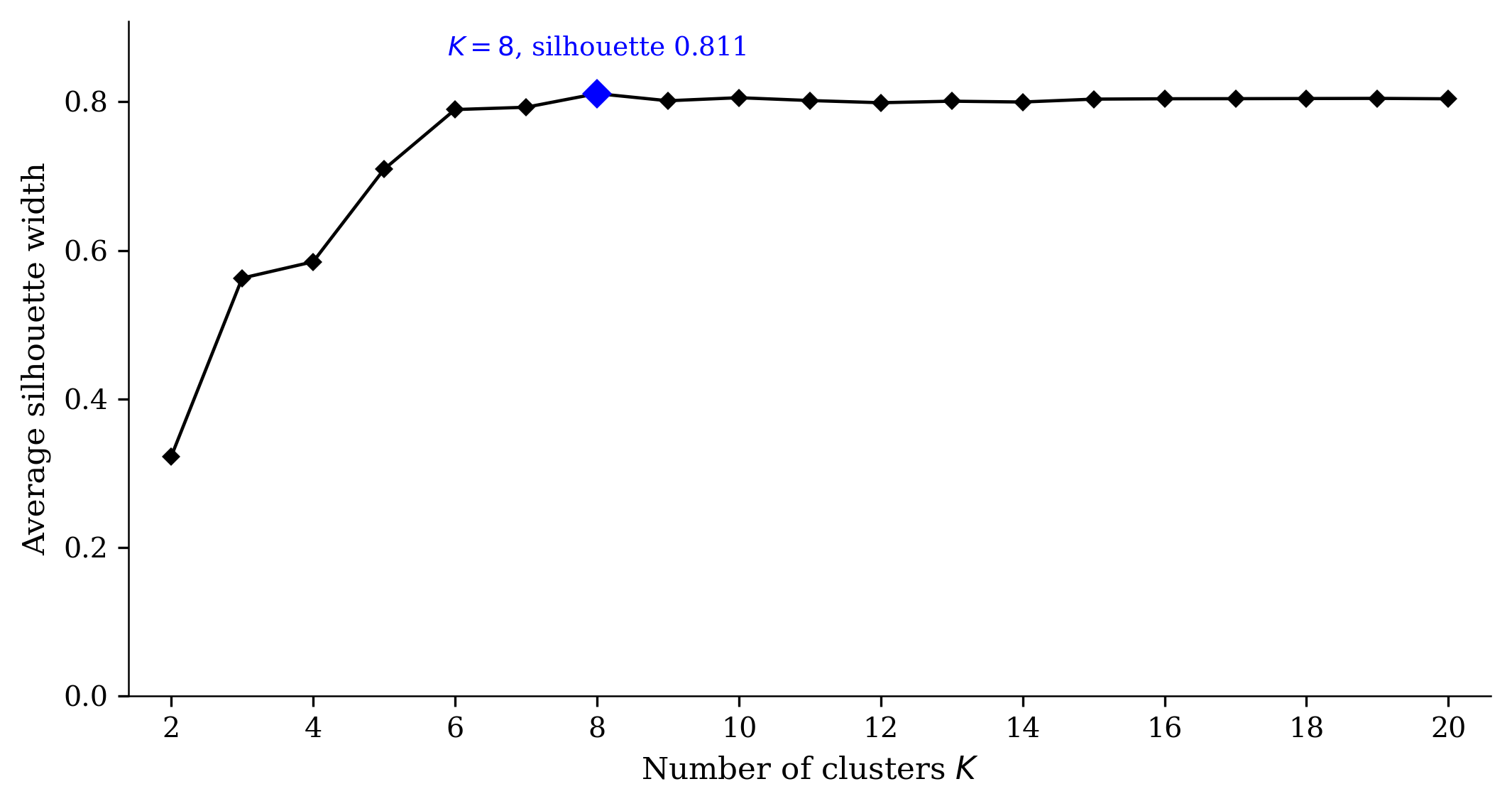}
\caption[Average silhouette width by number of clusters]{%
{Average silhouette width by number of clusters.}
Average silhouette width for the hierarchical clustering of 1,357 firms' portfolio-share trajectories, computed on the dynamic time warping distance matrix. The clustering uses pure DTW with a Sakoe-Chiba band of radius two, path-length-normalised distances, and average linkage. The width rises steeply to $K=8$ and is flat thereafter; $K=8$ attains the maximum over the range shown and is the value used in
Tables~\ref{tab:dtw_crosstab} and~\ref{tab:dtw_robustness}.%
}
\label{fig:dtw_silhoutte}
\end{figure}

\section{The Profitability Gap}
\label{app:gap}

\subsection{Technology-Specific Profitability Trends}
\label{app:technology_trends}

\begin{table}[H]
\centering
\caption{Technology-Specific ROA Trends in Firm-Year Data}
\label{tab:technology_trends}
\begin{threeparttable}
\footnotesize
\begin{tabular}{@{}lcccrr@{}}
\toprule
 & Pooled & Controls and & Firm fixed & Firm-years & Firms \\
 & firm-years & country FE & effects & & \\
 & (1) & (2) & (3) & & \\
\midrule
Wind & 0.366*** & 0.325*** & 0.369*** & 4,222 & 549 \\
 & (0.046) & (0.044) & (0.059) & & \\
Solar PV & 0.172 & 0.213* & 0.150 & 859 & 128 \\
 & (0.123) & (0.119) & (0.131) & & \\
Hydro & 0.070 & 0.036 & 0.090* & 2,306 & 245 \\
 & (0.047) & (0.045) & (0.053) & & \\
Nuclear & 0.048 & 0.026 & $-$0.073 & 211 & 19 \\
 & (0.106) & (0.084) & (0.073) & & \\
Gas & $-$0.033 & $-$0.026 & $-$0.011 & 2,596 & 265 \\
 & (0.041) & (0.039) & (0.046) & & \\
Coal & $-$0.105** & $-$0.078* & $-$0.083* & 452 & 42 \\
 & (0.045) & (0.044) & (0.049) & & \\
Oil & 0.031 & 0.018 & 0.117 & 362 & 48 \\
 & (0.092) & (0.099) & (0.129) & & \\
Biomass & $-$0.031 & $-$0.002 & 0.110 & 960 & 141 \\
 & (0.070) & (0.082) & (0.096) & & \\
\bottomrule
\end{tabular}

\begin{tablenotes}[flushleft]\footnotesize
\item \textit{Notes:} Each entry is the coefficient on a linear calendar-time trend
from a separate regression for the indicated dominant-technology group. ROA and the
coefficients are measured in percentage points, so estimates give the annual change in
ROA in percentage points. Column (1) uses all firm-year observations. Column (2) adds
lagged sales growth, firm size, leverage, and country fixed effects. Column (3) includes
firm fixed effects and is identified from firms observed at least twice within the
technology group. Firm-clustered standard errors are in parentheses. Firm-years and
firms refer to the pooled sample.
*** $p<0.01$, ** $p<0.05$, * $p<0.10$.
\end{tablenotes}
\end{threeparttable}
\end{table}

\subsection{Robustness of the Profitability Gap}
\label{app:gap_robustness}

Table~\ref{tab:gap_rob_overview} collects the headline estimate from each check;
the tables that follow report them in full. The widening stays positive and
significant at the 1 percent level throughout, ranging from 2.450 to 3.632
percentage points.

\begin{table}[!htbp]
\centering
\caption{Overview of Robustness Checks for the Post-2018 Widening}
\label{tab:gap_rob_overview}
\begin{threeparttable}
\footnotesize
\begin{tabular}{@{}lrrrr@{}}
\toprule
 & Estimate & Standard error & Firms & Firm-years \\
\midrule
Baseline & 3.227*** & 0.625 & 986 & 8,359 \\
\addlinespace
Break year 2018 & 2.897*** & 0.607 & 986 & 8,359 \\
Break year 2020 & 3.595*** & 0.667 & 986 & 8,359 \\
ROE instead of ROA & 5.146** & 2.450 & 965 & 8,000 \\
Parent companies excluded & 3.632*** & 0.735 & 875 & 7,039 \\
Subsidiaries excluded & 2.896*** & 0.660 & 784 & 6,852 \\
Single-unit firms excluded & 2.999*** & 0.643 & 714 & 6,498 \\
\bottomrule
\end{tabular}
\begin{tablenotes}[flushleft]\footnotesize
\item \textit{Notes:} Each row reports the post-break change in the Renewable--Fossil gap. The ROE row is measured on a larger scale and is not comparable in levels. Subsequent tables report each check in full. *** $p<0.01$.
\end{tablenotes}
\end{threeparttable}
\end{table}

\begin{table}[!htbp]
\centering
\caption{Robustness: alternative break years}
\label{tab:gap_rob_break}
\begin{threeparttable}
\footnotesize
\begin{tabular}{@{}lccc@{}}
\toprule
 & 2018 & 2019 & 2020 \\
 & (1) & (2) & (3) \\
\midrule
Change after 2018 & 2.897*** & 3.227*** & 3.595*** \\
 & (0.607) & (0.625) & (0.667) \\
\midrule
Firms & 986 & 986 & 986 \\
Firm-years & 8,359 & 8,359 & 8,359 \\
\bottomrule
\end{tabular}
\begin{tablenotes}[flushleft]\footnotesize
\item \textit{Notes:} Each column moves the year from which the post-break indicator switches on. Column (2) is the baseline. ROA in percentage points; country-by-year fixed effects and lagged firm controls throughout. *** $p<0.01$.
\end{tablenotes}
\end{threeparttable}
\end{table}

\begin{table}[!htbp]
\centering
\caption{Robustness: ROE instead of ROA}
\label{tab:gap_rob_roe}
\begin{threeparttable}
\footnotesize
\begin{tabular}{@{}lcc@{}}
\toprule
 & ROA & ROE \\
\midrule
Change after 2018 & 3.227*** & 5.146** \\
 & (0.625) & (2.450) \\
\midrule
Firms & 986 & 965 \\
Firm-years & 8,359 & 8,000 \\
\bottomrule
\end{tabular}
\begin{tablenotes}[flushleft]\footnotesize
\item \textit{Notes:} Column (2) replaces the outcome with return on equity, which is measured on a larger scale; the smaller sample reflects the positive-equity requirement. Outcome in percentage points; country-by-year fixed effects and lagged firm controls throughout. *** $p<0.01$.
\end{tablenotes}
\end{threeparttable}
\end{table}


\begin{table}[!htbp]
\centering
\caption{Robustness: parent companies excluded}
\label{tab:gap_rob_parents}
\begin{threeparttable}
\footnotesize
\begin{tabular}{@{}lcc@{}}
\toprule
 & Baseline & Parents excluded \\
 & (1) & (2) \\
\midrule
Change after 2018 & 3.227*** & 3.632*** \\
 & (0.625) & (0.735) \\
\midrule
Firms & 986 & 875 \\
Firm-years & 8,359 & 7,039 \\
\bottomrule
\end{tabular}
\begin{tablenotes}[flushleft]\footnotesize
\item \textit{Notes:} Column (2) drops firms recorded as the parent of another sample firm. ROA in percentage points; country-by-year fixed effects and lagged firm controls throughout. *** $p<0.01$.
\end{tablenotes}
\end{threeparttable}
\end{table}

\begin{table}[!htbp]
\centering
\caption{Robustness: subsidiaries excluded}
\label{tab:gap_rob_subs}
\begin{threeparttable}
\footnotesize
\begin{tabular}{@{}lcc@{}}
\toprule
 & Baseline & Subsidiaries excluded \\
 & (1) & (2) \\
\midrule
Change after 2018 & 3.227*** & 2.896*** \\
 & (0.625) & (0.660) \\
\midrule
Firms & 986 & 784 \\
Firm-years & 8,359 & 6,852 \\
\bottomrule
\end{tabular}
\begin{tablenotes}[flushleft]\footnotesize
\item \textit{Notes:} Column (2) drops firms whose recorded parent is also in the sample. ROA in percentage points; country-by-year fixed effects and lagged firm controls throughout. *** $p<0.01$.
\end{tablenotes}
\end{threeparttable}
\end{table}

\begin{table}[!htbp]
\centering
\caption{Robustness: single-unit firms excluded}
\label{tab:gap_rob_single}
\begin{threeparttable}
\footnotesize
\begin{tabular}{@{}lcc@{}}
\toprule
 & Baseline & Single-unit excluded \\
 & (1) & (2) \\
\midrule
Change after 2018 & 3.227*** & 2.999*** \\
 & (0.625) & (0.643) \\
\midrule
Firms & 986 & 714 \\
Firm-years & 8,359 & 6,498 \\
\bottomrule
\end{tabular}
\begin{tablenotes}[flushleft]\footnotesize
\item \textit{Notes:} Column (2) drops firms whose entire capacity consists of one generating unit. ROA in percentage points; country-by-year fixed effects and lagged firm controls throughout. *** $p<0.01$.
\end{tablenotes}
\end{threeparttable}
\end{table}

\begin{figure}[H]
\centering
\includegraphics[width=0.85\textwidth]{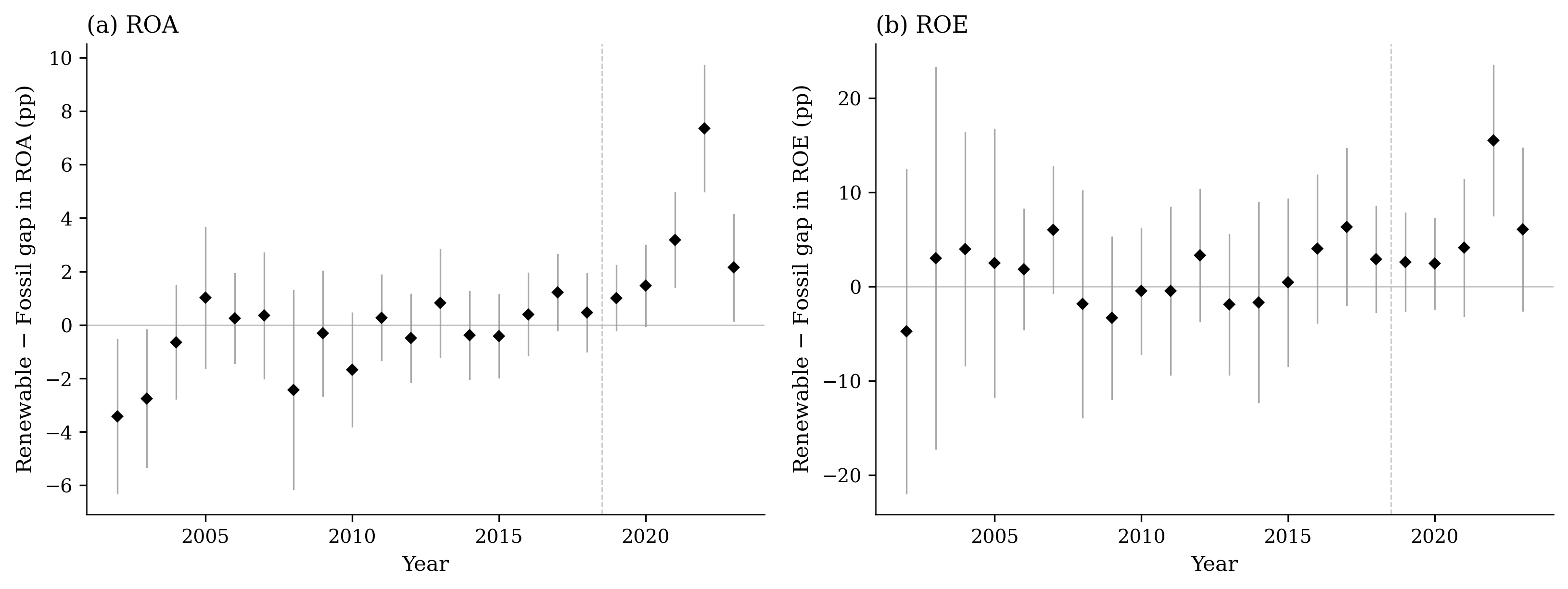}
\caption[The profitability gap measured by ROA and by ROE]{%
{The profitability gap measured by ROA and by ROE.}
Annual conditional Renewable--Fossil gaps from equation~\eqref{eq:gap}, with ROA
in panel~(a) and ROE in panel~(b). Both use country-by-year fixed effects, lagged
firm controls, and firm-clustered standard errors. The panels are on different
vertical scales: the ROE gap is roughly five times as large and roughly five
times as noisy. Its confidence intervals contain zero in every year except 2022,
so panel~(b) is consistent with panel~(a) rather than separate evidence on the
timing of the break.%
}
\label{app_fig:ROA_ROE_GAP}
\end{figure}

Table~\ref{tab:gap_rob_break} moves the break year; Table~\ref{tab:gap_rob_roe}
replaces ROA with ROE; Tables~\ref{tab:gap_rob_parents}--\ref{tab:gap_rob_single}
drop parent companies, subsidiaries, and single-unit firms in turn.
Figure~\ref{app_fig:ROA_ROE_GAP} plots the annual ROA and ROE gaps side by side.

Table~\ref{tab:gap_robustness} adds firm fixed effects and then restricts the
panel to firms observed throughout. Both reduce the estimate, to 2.558 and 2.450
percentage points, while leaving it significant at the 1 percent level. With firm
fixed effects the level of the gap is absorbed and identification comes only from
firms that change, so these are lower bounds on a difference that is largely
cross-sectional.

\begin{table}[!htbp]
\centering
\caption{Robustness of the Post-2018 Widening}
\label{tab:gap_robustness}
\begin{threeparttable}
\footnotesize
\begin{tabular}{@{}lccc@{}}
\toprule
 & Baseline & Firm FE & Constant sample \\
 & (1) & (2) & (3) \\
\midrule
Change after 2018        & 3.227*** & 2.558*** & 2.450*** \\
                         & (0.625) & (0.620) & (0.675) \\
\midrule
Country $\times$ year FE & Yes & Yes & Yes \\
Firm FE                  & No & Yes & Yes \\
Lagged firm controls     & Yes & Yes & Yes \\
Constant firm sample     & No & No & Yes \\
Firms                    & 986 & 922 & 448 \\
Firm-years               & 8,359 & 8,294 & 5,711 \\
\bottomrule
\end{tabular}

\begin{tablenotes}[flushleft]\footnotesize
\item \textit{Notes:}
The dependent variable is ROA in percentage points. Column (1) repeats the
change from Panel A of Table~\ref{tab:gap_prepost}, identified between firms
within a country-year. Columns (2) and (3) add firm fixed effects, so the
change is identified within firms over time; the level of the gap is then
absorbed and cannot be reported. Column (3) further restricts the sample to
firms first observed by 2015 and still observed in 2022, so entry and exit
around the 2019 break cannot drive the result. All columns include
country-by-year fixed effects and lagged firm controls. Standard errors are
clustered by firm.
*** $p<0.01$, ** $p<0.05$, * $p<0.10$.
\end{tablenotes}
\end{threeparttable}
\end{table}

Table~\ref{tab:winsorization} reports what the 1st/99th-percentile winsorization
does and re-estimates the gap without it. The change after 2018 is 3.067
percentage points against a baseline of 3.227, with the standard error doubling
to 1.240.

\begin{table}[!htbp]
\centering
\caption{Winsorization Diagnostics and the Post-2018 Profitability Gap}
\label{tab:winsorization}
\begin{threeparttable}
\footnotesize
\begin{tabular}{@{}lrrrrrr@{}}
\toprule
\multicolumn{7}{@{}l}{\textit{Panel A: Distributional effects in the four-industry sample}} \\
Variable & Raw mean & Raw SD & Wins. mean & Wins. SD & Bounds & Capped \\
\midrule
ROA & 3.75 & 28.84 & 4.04 & 9.38 & [$-$40.88, 51.22] & 150 (1.3\%) \\
ROE & 4.52 & 413.93 & 10.66 & 38.56 & [$-$238.78, 204.21] & 156 (1.4\%) \\
Sales growth, $t-1$ & 8.34 & 72.33 & 6.95 & 45.92 & [$-$213.38, 210.57] & 215 (1.8\%) \\
Leverage, $t-1$ & 32.30 & 66.20 & 30.96 & 25.05 & [0.00, 109.68] & 104 (0.9\%) \\
\midrule
\multicolumn{7}{@{}l}{\textit{Panel B: Gap estimates under alternative winsorization choices}} \\
Specification & \multicolumn{2}{c}{Change after 2018} & \multicolumn{2}{c}{Post-2018 gap} & Firms & Firm-years \\
 & Estimate & SE & Estimate & SE & & \\
\midrule
Baseline & 3.227*** & (0.625) & 3.195*** & (0.572) & 986 & 8,359 \\
No winsorization & 3.067** & (1.240) & 5.482*** & (1.232) & 986 & 8,359 \\
\bottomrule
\end{tabular}
\begin{tablenotes}[flushleft]\footnotesize
\item \textit{Notes:} Panel A compares raw and winsorized variables on the exact four-industry base sample. Consistent with the preprocessing procedure, the 1st/99th-percentile cutoffs are computed on the full panel before the industry and year restrictions. ROA, ROE, sales growth, leverage, and the reported bounds are in percentage points. Capped reports the number and share of non-missing observations changed by winsorization. Panel B re-estimates the pooled pre/post specification on a common Renewable--Fossil sample. The baseline winsorizes ROA, lagged sales growth, and lagged leverage; the alternative leaves all three variables unwinsorized. Firm size is never winsorized. Country-by-year fixed effects and firm-clustered standard errors throughout. $^{*}p<0.10$, $^{**}p<0.05$, $^{***}p<0.01$.
\end{tablenotes}
\end{threeparttable}
\end{table}

\newpage

\section{Price Exposure}
\label{app:price_exposure}

\subsection{Decomposing the Gap into Price Components}
\label{subsec:decomposition}

Equation~\eqref{eq:mechanism} estimates how firms' carbon and gas exposure are associated with profitability as the corresponding prices change.
We translate these coefficients into model-implied components of the
Renewable--Fossil profitability gap.

Let $\widehat{\theta}_{t}$ denote the annual Renewable--Fossil profitability
gap estimated from equation~\eqref{eq:gap}. We express each annual gap
relative to its mean over the pre-2019 years of the mechanism estimation
window, $\mathcal{P}=\{2008,\ldots,2018\}$:

\begin{equation}
\Delta\widehat{\theta}_{t}
=
\widehat{\theta}_{t}
-
\frac{1}{|\mathcal{P}|}
\sum_{s\in\mathcal{P}}
\widehat{\theta}_{s}.
\label{eq:gap_change}
\end{equation}

The series $\Delta\widehat{\theta}_{t}$ measures the annual change in the
Renewable--Fossil gap relative to its average before 2019. It is zero on
average over 2008--2018 by construction.

We define the mean exposure differences between Fossil- and
Renewable-archetype firms as

\begin{equation}
\Delta \mathrm{CI}
=
\overline{\mathrm{CI}}^{\,F}
-
\overline{\mathrm{CI}}^{\,R},
\qquad
\Delta s^{\mathrm{gas}}
=
\overline{s}^{\,\mathrm{gas},F}
-
\overline{s}^{\,\mathrm{gas},R},
\label{eq:exposure_gaps}
\end{equation}

where the bars denote sample means in the mechanism estimation sample. The
exposure differences are defined as Fossil minus Renewable, whereas the
profitability gap is defined as Renewable minus Fossil.

The model-implied carbon-price component in year $t$ is

\begin{equation}
C^{\mathrm{CP}}_{t}
=
-
\widehat{\beta}_{1}
\Delta \mathrm{CI}
\left(
\widetilde{\mathrm{CP}}_{t}
-
\frac{1}{|\mathcal{P}|}
\sum_{s\in\mathcal{P}}
\widetilde{\mathrm{CP}}_{s}
\right).
\label{eq:carbon_component}
\end{equation}

The leading minus sign reconciles the opposite ordering of the profitability
and exposure differences. When $\widehat{\beta}_{1}<0$ and
Fossil-archetype firms have greater carbon intensity, an increase in the
allowance price produces a positive carbon-price component of the
Renewable--Fossil gap.

The gas-price component is defined analogously:

\begin{equation}
C^{\mathrm{GP}}_{t}
=
-
\widehat{\beta}_{2}
\Delta s^{\mathrm{gas}}
\left(
\widetilde{\mathrm{GP}}_{t}
-
\frac{1}{|\mathcal{P}|}
\sum_{s\in\mathcal{P}}
\widetilde{\mathrm{GP}}_{s}
\right).
\label{eq:gas_component}
\end{equation}

This yields three annual series: the observed change in the gap,
$\Delta\widehat{\theta}_{t}$, the carbon-price component,
$C^{\mathrm{CP}}_{t}$, and the combined carbon- and gas-price component,
$C^{\mathrm{CP}}_{t}+C^{\mathrm{GP}}_{t}$. All three are measured relative to the same
2008--2018 mean. The comparison therefore concerns changes in the gap rather
than its level and does not require a separately estimated intercept.

We then average the observed change and the two price
components over the post-2018 period,
$\mathcal{Q}=\{2019,\ldots,2023\}$:

\begin{equation}
\overline{\Delta\theta}
=
\frac{1}{|\mathcal{Q}|}
\sum_{t\in\mathcal{Q}}
\Delta\widehat{\theta}_{t},
\qquad
\overline{C}^{CP}
=
\frac{1}{|\mathcal{Q}|}
\sum_{t\in\mathcal{Q}}
C^{\mathrm{CP}}_{t},
\qquad
\overline{C}^{GP}
=
\frac{1}{|\mathcal{Q}|}
\sum_{t\in\mathcal{Q}}
C^{\mathrm{GP}}_{t}.
\label{eq:average_components}
\end{equation}

We define the part not captured by the two price-exposure terms as

\begin{equation}
R
=
\overline{\Delta\theta}
-
\overline{C}^{CP}
-
\overline{C}^{GP}.
\label{eq:remainder}
\end{equation}

The decomposition reported in Section~\ref{subsec:carbonprice_results} is therefore the
accounting identity

\begin{equation}
\overline{\Delta\theta}
=
\overline{C}^{CP}
+
\overline{C}^{GP}
+
R.
\label{eq:gap_decomposition}
\end{equation}

The three terms are model-implied accounting components rather than
identified counterfactual effects. Their interpretation assumes linear and constant price--exposure relationships over the sample period. Table~\ref{tab:price_exposure_accounting} reports the
inputs and results of the decomposition: Panel~A gives the archetype exposure
differences and the implied ROA change per reporting increment, Panel~B the
average components over 2019--2023 with standard errors. The underlying exposure differences between the two archetypes are highly persistent over time, as shown in Appendix Figure~\ref{app_fig:exposure_by_archetype}.

\begin{table}[!t]
\centering
\caption{Exposure Differences and Price-Exposure Accounting Components}
\label{tab:price_exposure_accounting}
\begin{threeparttable}
\footnotesize
\begin{tabular}{@{}lrrr@{}}
\toprule
\multicolumn{4}{@{}l}{\textit{Panel A: Exposure and scaling inputs}} \\
Exposure & Archetype difference & Reporting increment & Implied ROA change \\
\midrule
Carbon intensity & 0.4787 tCO$_2$/MWh & 0.1 tCO$_2$/MWh & -0.176 pp \\
Gas-capacity share & 0.6821 & 0.1 (10 pp) & -0.137 pp \\
\addlinespace
\multicolumn{4}{@{}l}{\textit{Panel B: Average change over 2019--2023}} \\
Component & Percentage points & Standard error & Share of widening \\
\midrule
Observed widening & 3.262 & (0.637) & 100.0\% \\
Carbon-price component & 1.294 & (0.285) & 39.7\% \\
Gas-price component & 0.607 & (0.107) & 18.6\% \\
Carbon- and gas-price components & 1.901 & (0.255) & 58.3\% \\
Not captured by the price terms & 1.362 & (0.686) & 41.7\% \\
\bottomrule
\end{tabular}
\begin{tablenotes}[flushleft]\footnotesize
\item \textit{Notes:} Panel A uses the 10,888 firm-years in the common complete-case sample. Archetype differences are Fossil minus Renewable. The final column scales each joint coefficient in column~(3) of Table~\ref{tab:carbon_intensity} to a 0.1-unit exposure difference and a one-standard-deviation increase in the corresponding price. Panel B reports the accounting components defined in equations~\eqref{eq:average_components} and~\eqref{eq:remainder}. The observed widening is the mean annual Renewable--Fossil gap over 2019--2023 relative to its 2008--2018 mean. Carbon and gas components scale the joint coefficients by the exposure differences in Panel A and the corresponding standardized price movements. Standard errors for the observed widening are clustered by firm; component standard errors use the delta method and the year-clustered mechanism covariance matrix. The residual standard error adds the widening and joint-component variances and therefore assumes zero covariance across the two estimations. Components are accounting quantities, not causal shares.
\end{tablenotes}
\end{threeparttable}
\end{table}

\begin{figure}[H]
\centering
\includegraphics[width=0.85\textwidth]{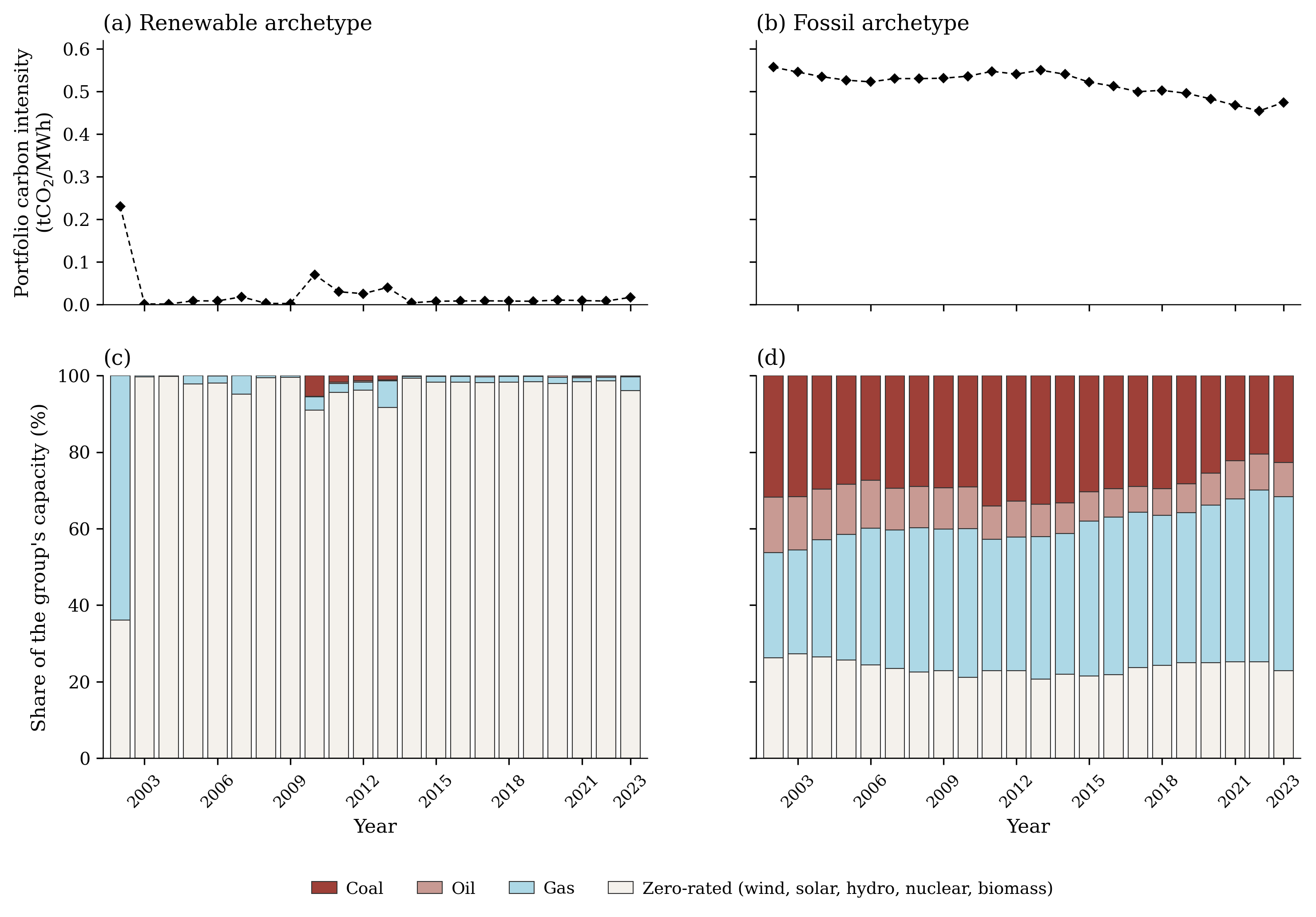}
\caption[Carbon Intensity by Portfolio Archetypes]{%
{Carbon Intensity by Portfolio Archetypes.}
Panels~(a) and~(b) plot each archetype's capacity-weighted portfolio carbon intensity; panels~(c) and~(d) show the composition behind it, with wind, solar, hydro, nuclear and biomass pooled as zero-rated because none carries direct combustion emissions. The exposure gap between the two archetypes stays between 0.47 and 0.55~tCO$_2$/MWh across the whole period, which is what allows the linear decomposition in Appendix~\ref{subsec:decomposition} to use its period average. The fossil group's intensity falls from 0.56 in 2002 to 0.48 in 2023 as coal gives way to gas, but does so gradually and without a break at 2019.%
}
\label{app_fig:exposure_by_archetype}
\end{figure}

\newpage

\subsection{Robustness of the Price-Exposure Estimates}
\label{app:carbonprice_robustness}

This subsection reports additional specifications for the relationship between
portfolio carbon intensity, the EU ETS allowance price, and firm profitability.
The analyses test the functional form imposed on the price terms, then vary the
inference procedure and the construction of portfolio carbon intensity.

Table~\ref{tab:flexible_tests} tests equation~\eqref{eq:mechanism}, which forces
each exposure coefficient to move linearly with its price, against the flexible
specification in equation~\eqref{eq:flexible_exposure}, which lets every year
have its own. The post-2018 deviations are jointly different from zero for both
exposures, but the linear restriction is not rejected.

\begin{table}[!htbp]
\centering
\caption{Tests of the Flexible Annual Price-Exposure Specification}
\label{tab:flexible_tests}
\begin{threeparttable}
\footnotesize
\begin{tabular}{@{}lccc@{}}
\toprule
Null hypothesis & Carbon intensity & Gas share & Joint \\
\midrule
Post-2018 deviations jointly zero & 16.87 [0.005] & 13.82 [0.017] & --- \\
Linear price restriction & 16.88 [0.262] & 13.50 [0.488] & 32.59 [0.251] \\
\bottomrule
\end{tabular}
\begin{tablenotes}[flushleft]\footnotesize
\item \textit{Notes:} Wald statistics with $p$-values in brackets, from the flexible specification in equation~\eqref{eq:flexible_exposure}. The first row tests whether the five post-2018 annual coefficients differ jointly from their 2008--2018 mean; it has five degrees of freedom. The second row tests the restriction that each exposure coefficient moves linearly with its price, as equation~\eqref{eq:mechanism} imposes; the separate tests have 14 degrees of freedom and the joint test has 28. Inference clusters by firm, because the two-way covariance matrix for the full set of annual coefficients is not positive semidefinite.
\end{tablenotes}
\end{threeparttable}
\end{table}

Table~\ref{tab:flexible_decomposition} re-runs the decomposition on the
year-specific coefficients instead. The carbon-price component rises from 1.294
to 1.897 percentage points, or from 40 to 58 percent of the widening, so the
paper's reported component is the more conservative of the two.

\begin{table}[!htbp]
\centering
\caption{The Carbon-Price Component without the Linearity Assumption}
\label{tab:flexible_decomposition}
\begin{threeparttable}
\footnotesize
\begin{tabular}{@{}lcc@{}}
\toprule
 & Linear in the price & Flexible by year \\
 & (1) & (2) \\
\midrule
Carbon-price component & 1.294 & 1.897 \\
 & (0.285) & (0.456) \\
Share of the widening & 40\% & 58\% \\
\midrule
Exposure gap $\Delta CI$ & 0.4787 & 0.4787 \\
Firm-years & 10,888 & 10,888 \\
\bottomrule
\end{tabular}

\begin{tablenotes}[flushleft]\footnotesize
\item \textit{Notes:} Column (1) is the paper's component, which scales a single
interaction coefficient by the standardized allowance-price deviation and therefore
assumes the coefficient moves linearly with the price. Column (2) replaces the price term
with the year-specific coefficient estimated from a full set of year interactions, so no
functional form for the price is imposed. Both scale by the same exposure gap and use
the same sample, firm and year fixed effects, and controls. Column (1) uses the
year-clustered mechanism covariance matrix. Column (2) uses firm-clustered
inference because the two-way covariance matrix for all annual coefficients is
indefinite; clustering does not change the point estimates.
\end{tablenotes}
\end{threeparttable}
\end{table}

\begin{figure}[!t]
\centering
\includegraphics[width=\textwidth]{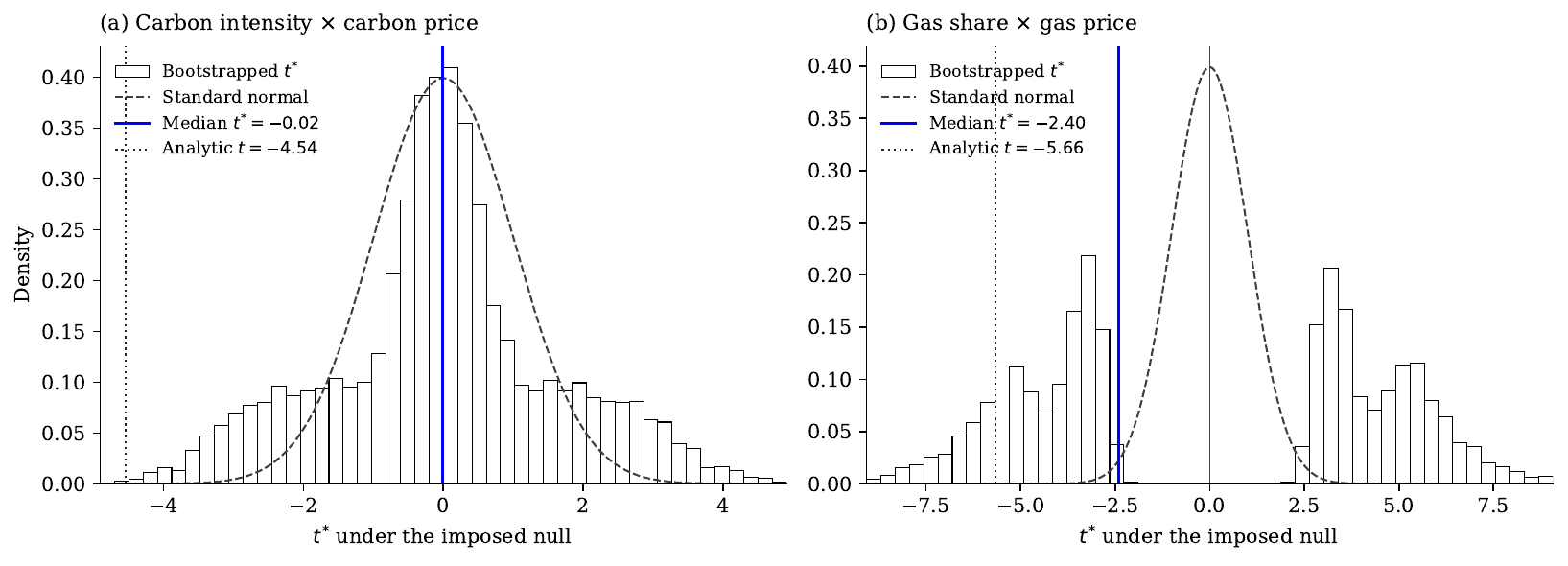}
\caption[Wild cluster bootstrap null distributions]{%
{Wild cluster bootstrap null distributions.}
Distribution of the bootstrapped $t$-statistic under the imposed null, from
9,999 replications with Rademacher weights on 16 year clusters, for the two
interactions of column~(3) of Table~\ref{tab:carbon_intensity}. Both panels come
from the same regression. The dashed curve is the standard normal, drawn in both
panels as a common yardstick; the horizontal scales differ because the gas
distribution is roughly three times as wide. The solid vertical line marks the
median bootstrapped $t$ and the dotted line the analytic $t$ against which the
bootstrap $p$-value is computed.

}
\label{app_fig:bootstrap_null}
\end{figure}

\begin{table}[!htbp]
\centering
\caption{Wild Cluster Bootstrap Diagnostic}
\label{tab:bootstrap_diagnostic}
\begin{threeparttable}
\footnotesize
\begin{tabular}{@{}llrrrl@{}}
\toprule
Specification & Interaction & Median $t^{*}$ & S.d. $t^{*}$ & $p$ & Reported \\
\midrule
(1) gas only & Gas share $\times$ GP & $-$2.58 & 6.63 & 0.2238 & No \\
(2) carbon only & Carbon intensity $\times$ CP & $-$0.10 & 1.44 & 0.0002 & Yes \\
(3) both & Carbon intensity $\times$ CP & $-$0.02 & 1.63 & 0.0026 & Yes \\
(3) both & Gas share $\times$ GP & $-$2.40 & 4.76 & 0.2172 & No \\
\bottomrule
\end{tabular}

\begin{tablenotes}[flushleft]\footnotesize
\item \textit{Notes:}
Moments of the bootstrapped $t$-statistic under the imposed null, from 9,999
replications with Rademacher weights on 16 year clusters.
\end{tablenotes}
\end{threeparttable}
\end{table}

Table~\ref{tab:alt_exposure} replaces portfolio carbon intensity with the
fossil and renewable capacity shares, which require none of the efficiency
assumptions in Table~\ref{tab:assumptions}.

\begin{table}[!htbp]
\centering
\caption{Robustness: capacity shares instead of the intensity index}
\label{tab:alt_exposure}
\begin{threeparttable}
\footnotesize
\begin{tabular}{@{}lcc@{}}
\toprule
 & Carbon intensity & Capacity shares \\
 & (1) & (2) \\
\midrule
Carbon intensity $\times$ CP & $-$1.756*** & --- \\
 & (0.457) &  \\
\addlinespace
Fossil share $\times$ CP & --- & $-$0.179 \\
 &  & (0.391) \\
\addlinespace
Renewable share $\times$ CP & --- & 1.194* \\
 &  & (0.662) \\
\addlinespace
\quad Renewable $-$ fossil & --- & 1.373** \\
 &  & (0.609) \\
\addlinespace
Gas share $\times$ GP & $-$1.371*** & $-$1.215*** \\
 & (0.296) & (0.336) \\
\midrule
Firms & 1,265 & 1,265 \\
Firm-years & 10,888 & 10,888 \\
Within $R^2$ & 0.0271 & 0.0296 \\
\bottomrule
\end{tabular}
\begin{tablenotes}[flushleft]\footnotesize
\item \textit{Notes:} Column (2) replaces portfolio carbon intensity with the
fossil and renewable capacity shares, each interacted with the allowance price, on
the same observations. The two columns are not comparable in levels: carbon
intensity is measured in tonnes of CO$_2$ per MWh and the shares in units of share.
Both columns include firm and year
fixed effects and lagged controls, with standard errors clustered two-way on firm
and year. *** $p<0.01$, ** $p<0.05$, * $p<0.10$.
\end{tablenotes}
\end{threeparttable}
\end{table}

\end{document}